\documentclass[apj]{emulateapj}
\usepackage{graphicx}
\usepackage{epstopdf}
\usepackage{soul}
\usepackage{amsmath}
\usepackage{color}

\slugcomment{Accepted for publication in {\it The Astrophysical Journal}}

\begin{document}

\title{How Robust Are the Size Measurements of High-redshift Compact Galaxies?}

\author{Roozbeh Davari\altaffilmark{1,3},
Luis C. Ho\altaffilmark{2,3},
Chien Y. Peng\altaffilmark{4}, and
Song Huang\altaffilmark{5}
}
\altaffiltext{1}{University of California, Riverside 900 University Avenue, Riverside, CA 92521, USA}
\altaffiltext{2}{Kavli Institute for Astronomy and Astrophysics, Peking University, Beijing 100871, P. R. China}
\altaffiltext{3}{The Observatories of the Carnegie Institution for Science 813 Santa Barbara Street, Pasadena, CA 91101, USA}
\altaffiltext{4}{Giant Magellan Telescope Organization 251 South Lake Avenue, Suite 300 Pasadena, CA 91101, USA }
\altaffiltext{5}{School of Space Science and Astronomy, Nanjing University, Nanjing 210093, P. R. China}

\begin{abstract}
Massive quiescent galaxies at $z \approx 2$ are apparently much more compact than galaxies of comparable mass today.  How robust are these size measurements?  We perform comprehensive simulations to determine possible biases and uncertainties in fitting single-component light distributions to real galaxies. In particular, we examine the robustness of the measurements of the luminosity, size, and other structural parameters.  We devise simulations  with increasing realism to systematically disentangle effects due to the technique (specifically using {\tt GALFIT}) and the intrinsic structures of the galaxies.  By accurately capturing the detailed substructures of nearby elliptical galaxies and then rescaling their sizes and signal-to-noise to mimic galaxies at different redshifts, we confirm that the massive quiescent galaxies at $z \approx 2$ are significantly more compact intrinsically than their local counterparts. Their observed compactness is not a result of missing faint outer light due to systematic errors in modeling. In fact, we find that fitting multi-component galaxies with a single S\'ersic profile, the procedure most commonly adopted in the literature, biases the inferred sizes higher by up to 10\%--20\%, which accentuates the amount of size evolution required. If the sky estimation has been done robustly and the model for the point-spread function is fairly accurate, {\tt GALFIT} can retrieve the properties of single-component galaxies over a wide range of signal-to-noise ratios without introducing any systematic errors. 
\end{abstract}

\section{INTRODUCTION}

Galaxy morphology has always been an important observable for understanding the formation and evolution of galaxies. Recent imaging studies using the {\it Hubble Space Telescope (HST)}/Wide Field Camera 3 (WFC3) and Advanced Camera for Surveys (ACS) have broadened our understanding of the formation and evolution of galaxies. The advent of WFC3 has given access to the rest-frame optical light of galaxies at $z \approx 2$.  Morphological studies of these galaxies show that quiescent systems make up a considerable fraction of the massive galaxy population at $z \approx 2$ (e.g., \citealt{Franx03}; \citealt{Daddi05}; \citealt{Kriek06}). Their structural evolution has been the subject of considerable interest, focusing in particular on their extremely compact nature compared to low-redshift galaxies of similar mass (e.g., \citealt{Daddi05}; \citealt{Toft07}; \citealt{Trujillo07}; Buitrago et al. 2008; \citealt{Cimatti08}; \citealt{Franx08}; \citealt{vanderWel08}; \citealt{vanDokkum08}; \citealt{Damjanov09}; \citealt{Hopkins09}; \citealt{Cassata10, Cassata11}; \citealt{Mancini10}; \citealt{Newman12}; \citealt{Szomoru12}). The sizes of these ``red nuggets,'' less than $\sim 1$ kpc, seem comparable to and sometimes even smaller than the point-spread function (PSF). The early formation and subsequent size growth of these massive, compact objects present a challenge to current models of galaxy formation and evolution (e.g., \citealt{Wuyts10}; \citealt{Oser12}). It is not clear via what pathways they become the massive galaxies of today. The rarity of these massive, compact galaxies at low redshift implies considerable size evolution between \emph{z} = 2 and \emph{z} = 0 (\citealt{vanDokkum08}; \citealt{Trujillo09}; \citealt{Taylor10}; but see \citealt{Saracco10}; \citealt{Valentinuzzi10}; \citealt{Poggianti13}). On average, from \emph{z} $\approx 2$, these objects would have to increase their size by 3--4 times while doubling their stellar mass (\citealt{vanDokkum10}; see also \citealt{Ichikawa12}). However, efforts to accurately quantify this evolution are hindered by potential uncertainties in measurement techniques: the mass densities may simply be systematically overestimated due to modeling uncertainties of photometric masses, and/or the sizes may be underestimated due to a lack of imaging depth or other measurement issues (\citealt{Hopkins09}; \citealt{Muzzin09}). 

In addition to the basic question of how these high-\emph{z} galaxies evolve in size, there is also still much debate about how these systems evolve in terms of their fundamental morphological type. By performing {\tt GALFIT} fitting on a sample of 14 compact, massive galaxies at \emph{z} = 2, \citet{vanderWel11} find that a significant subset of their sample appears highly flattened in projection, which, considering viewing angle statistics, implies those galaxies have pronounced disks.  They claim that 65\%$\pm$15\% of the population of massive quiescent at \emph{z} $>$ 2 galaxies are disk-dominated. \citet{Bruce12} find that a considerable fraction (25\%$\pm$6\% using a definition of bulge-to-total ratio $B/T < 0.5$ and 40\%$\pm$7\% using a definition of $n < 2.5$ for disk-dominated) of the most quiescent galaxies, with specific star formation rates $< 10^{-10} \ {\rm yr}^{-1}$, have disk-dominated morphologies, including a small number of essentially pure disk galaxies (with $B/T <$ 0.1). They claim that these passive disks appear to be normal disks in the sense that they show an axial-ratio distribution comparable to that displayed by present-day disks. This implies that while the massive galaxy population is undergoing dramatic changes at this crucial epoch, the physical mechanisms that quench star formation activity are not obviously connected to those responsible for transforming the morphologies of massive \emph{z} $\approx$ 2 galaxies into present-day giant ellipticals.

A recent morphological study of nearby elliptical galaxies has opened a new window to understanding the formation and assembly of early-type galaxies.  \citet[][hereafter H13]{Huang13a} present a detailed, comprehensive structural analysis of about 100 representative, nearby elliptical galaxies spanning a range of environments and stellar masses ($M_*$ $\approx $ $10^{10.2}-10^{12.4}$ $M_{\Sun}$). They use {\tt GALFIT 3.0} (\citealt{Peng02, Peng10}) to perform two-dimensional, multi-component decomposition of relatively deep, moderately high-resolution $V$-band images acquired as a part of the Carnegie-Irvine Galaxy Survey\footnote{\tt http://cgs.obs.carnegiescience.edu/CGS/Home.html} (CGS; \citealt{Ho11}). H13 challenge the conventional notion that the main body of giant ellipticals follows a single structure described by a high S\'{e}rsic (1968) index (e.g., $n \geq 3 - 4$). They propose that the global light distribution of the majority ($>$75\%) of ellipticals are best described by three S\'{e}rsic components: a compact, inner core with typical effective radius $R_e$ $<$ 1 kpc and luminosity fraction $f \approx 0.1 - 0.15$; an intermediate-scale, middle component with $R_e \approx 2.5$ kpc and $f \approx 0.2 - 0.25$; and an extended, outer envelope with $R_e \approx 10$ kpc and $f \approx 0.6$. The subcomponents have relatively low S\'{e}rsic indices, in the range $n \approx 1-2$. They also find that the ellipticity of the isophotes systematically increases toward large radii. They believe that the combination of their model's inner and middle components for high-luminosity ellipticals resembles the compact, massive galaxies at high-\emph{z} (\citealt{Huang13b}).

Despite there being codes that can automate the measurement of galaxy properties (e.g., \citealt{Simard02, Simard11}; \citealt{Barden12}), there are still many disagreements over the analysis techniques, because they are often viewed as being too simple, and what the measurements mean.  Even stepping up the sophistication, however, makes interpretation non-trivial unless one has physical motivations for doing so (H13).  These disagreements inspire other techniques (\citealt{Conselice97}; \citealt{Lotz04}) to measure galaxies non-parametrically, although each one has its own benefits and shortcomings.  In addition to galaxy shapes being difficult to quantify, many other factors complicate the image analysis of galaxies, including determination of the sky background, the brightness of the galaxy [signal-to-noise ratio ($S/N$)], the resolution of the images, $(1+z)^4$ surface brightness dimming, finding the best PSF, the method employed for modeling the galaxy, and the potential biases of the fitting pipeline. Galaxy simulations are invaluable tools for understanding the performance of quantitative fitting pipelines because they provide control over the aforementioned factors (e.g., \citealt{Trujillo07}; \citealt{Cimatti08}; \citealt{Mancini10}; \citealt{Szomoru10,Szomoru12}; \citealt{vanDokkum10}; \citealt{Williams10}; \citealt{Papovich12}; and \citealt{vanderWel12} for high-\emph{z} galaxies and \citealt{Haussler07} and \citealt{Meert13} for low-\emph{z} galaxies). 

By taking advantage of existing observations, detailed analysis, and the multi-component picture of the local elliptical galaxies from H13, we address two key questions: 

1) Do the quiescent massive galaxies at \emph{z} = 2 have sizes comparable with the quiescent massive local galaxies and therefore their observed compactness is not intrinsic but an artifact of inappropriate and insufficient morphological modeling? 

2) What are the true uncertainties of the size and total luminosity measurements at different $S/N$ levels due to structural complexities? 

Two sets of simulations are performed to address these questions. Model galaxies with a single S\'{e}rsic component with parameter ranges that cover the {\it HST}/WFC3 observations of the red nuggets are the first set of simulations we perform. These idealized simulations can determine the robustness of fitting pipelines at different noise levels (\citealt{Haussler07}). Therefore, the uncertainties and systematic errors measured from these simulated galaxies can be considered as lower limits. More informative analysis is done by simulating multi-component galaxies with the properties obtained from H13. By artificially scaling modern-day galaxies to sizes and  luminosities comparable to those found in galaxies at $z \approx 2$, we can fit them with single-component S\'{e}rsic models to understand the systematics caused by complexities in galaxy structures. The differences between the input and output parameters shed light on possible uncertainties and biases in size and total luminosity measurements of the high-\emph{z} galaxies.

This paper is organized as follows. Section 2 gives the details of the galaxy simulations used throughout this study. In Section 3, we present the main results of our {\tt GALFIT} models. In Section 4, we compare our results with similar studies in the literature. Implications for red nuggets are discussed in Section 5, ending with a summary in Section 6. 

All the results assume a standard cosmology ($\emph{H}_0$ = 71 ${\rm km^{-1}\, s^{-1}\, Mpc^{-1}}$, $\Omega_m$ = 0.27, and $\Omega_{\Lambda}$ = 0.73) and AB magnitudes.

\section{SIMULATIONS}

Simulations can determine the robustness of fitting pipelines like {\tt GALFIT}\footnote{\tt http://users.obs.carnegiescience.edu/peng/work/galfit/galfit.html}. In this section, we describe the different sets of simulations in detail. 

Two sets of objects are simulated in order to address different issues.  The first, using a single-component S\'{e}rsic profile, is to establish the baseline capability due only to $S/N$, under the most idealized situations, and in the absence of any complexities.  This sets the fundamental limits of the technique.  The second set of simulations, by rescaling models of nearby elliptical galaxies and all their sub-components to comparable sizes and $S/N$ found at high \emph{z} and fitting them using a single-component S\'{e}rsic model, directly tests the null hypothesis that early-type galaxies have not evolved in size since \emph{z} $\approx$ 2.  We use Ned Wright's cosmology calculator\footnote{\tt http://www.astro.ucla.edu/$\sim$wright/CosmoCalc.html} to compute redshift-dependent quantities. 

\begin{figure*}[t]
  \centering
\includegraphics[width=140mm,angle=-90]{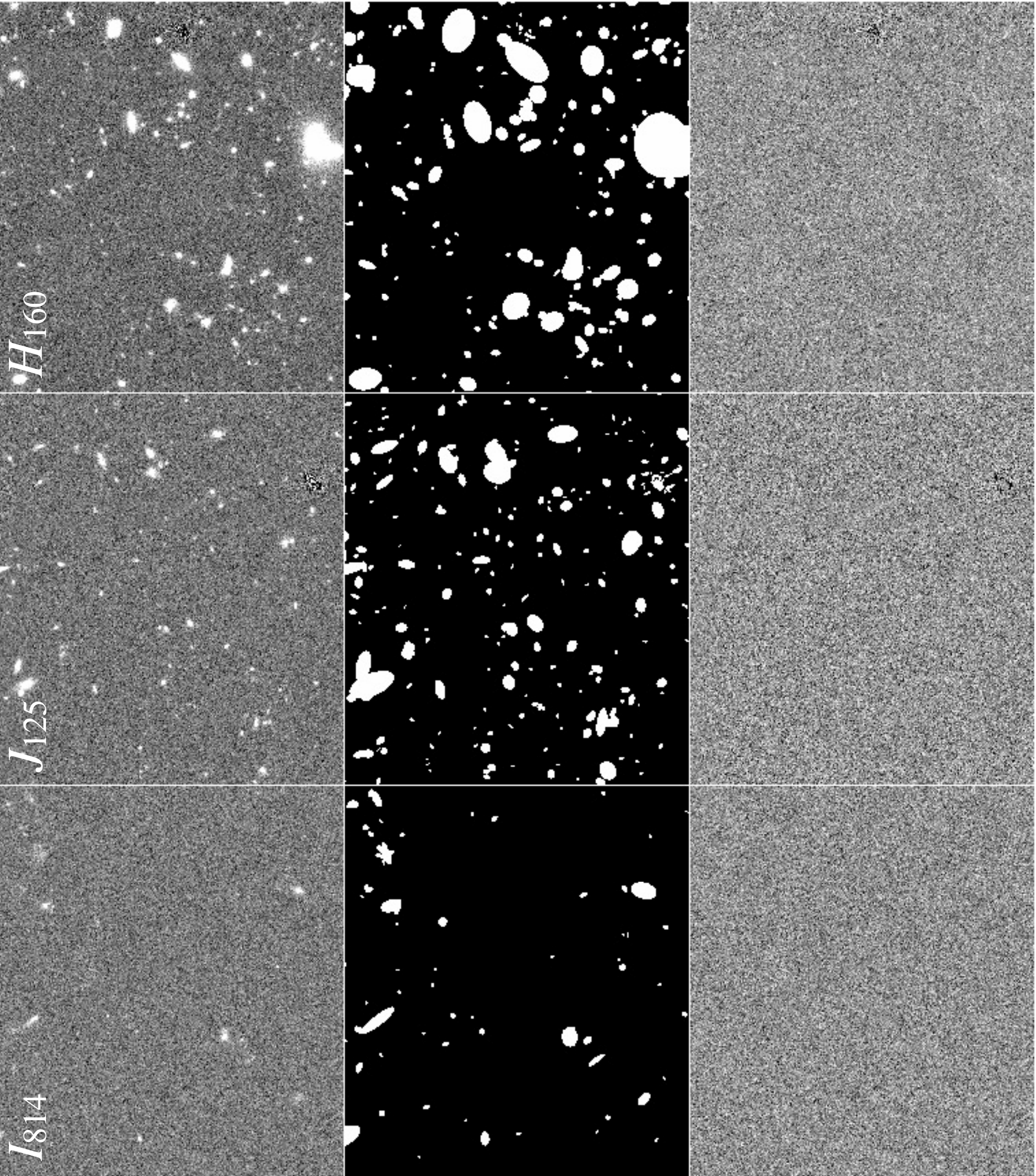}
  \caption{Sky backgrounds drawn from CANDELS UDS mosaic images in $I_{814}$, $J_{125}$, and $H_{160}$. The simulated galaxies are put at random positions on these backgrounds. The regions are 1021 $\times$ 1021 pixels (i.e., $\sim$30$\arcsec$ in $I_{814}$ and $\sim$60$\arcsec$ in $J_{125}$ and $H_{160}$), which is the size of the {\it HST}/ACS and {\it HST}/WFC3 CCDs. The middle panels show the bad pixel mask images of each sky background; the bottom panels are the backgrounds after removing the undesired objects. In the bottom panels, the masked pixels are replaced by randomly chosen background pixels. The bad pixel masks are used for {\tt GALFIT} modeling.\label{fig:skies}}
\end{figure*}

\subsection{Method}

We use {\tt GALFIT 3.0} for the simulations. {\tt GALFIT} is an image analysis algorithm that can model profiles of galaxies, stars, and other astronomical objects in digital images. If successful, the features of interest are summarized into a small set of numbers, such as size, luminosity, profile central concentration, and geometrical parameters. {\tt GALFIT} uses several common functions in the astronomical literature, including: exponential, S\'{e}rsic/de Vaucouleurs, Nuker, Gaussian, King, and Moffat. Out of all the functions, we exploit only the S\'{e}rsic (1968) profile and the sky background component. The sky component fits the background with a plane with a constant slope and therefore can correct for any non-flatness to first order. The S\'{e}rsic component describes the radial surface brightness profile of a galaxy as

\begin{figure*}[t]
  \centering
\includegraphics[width=181mm]{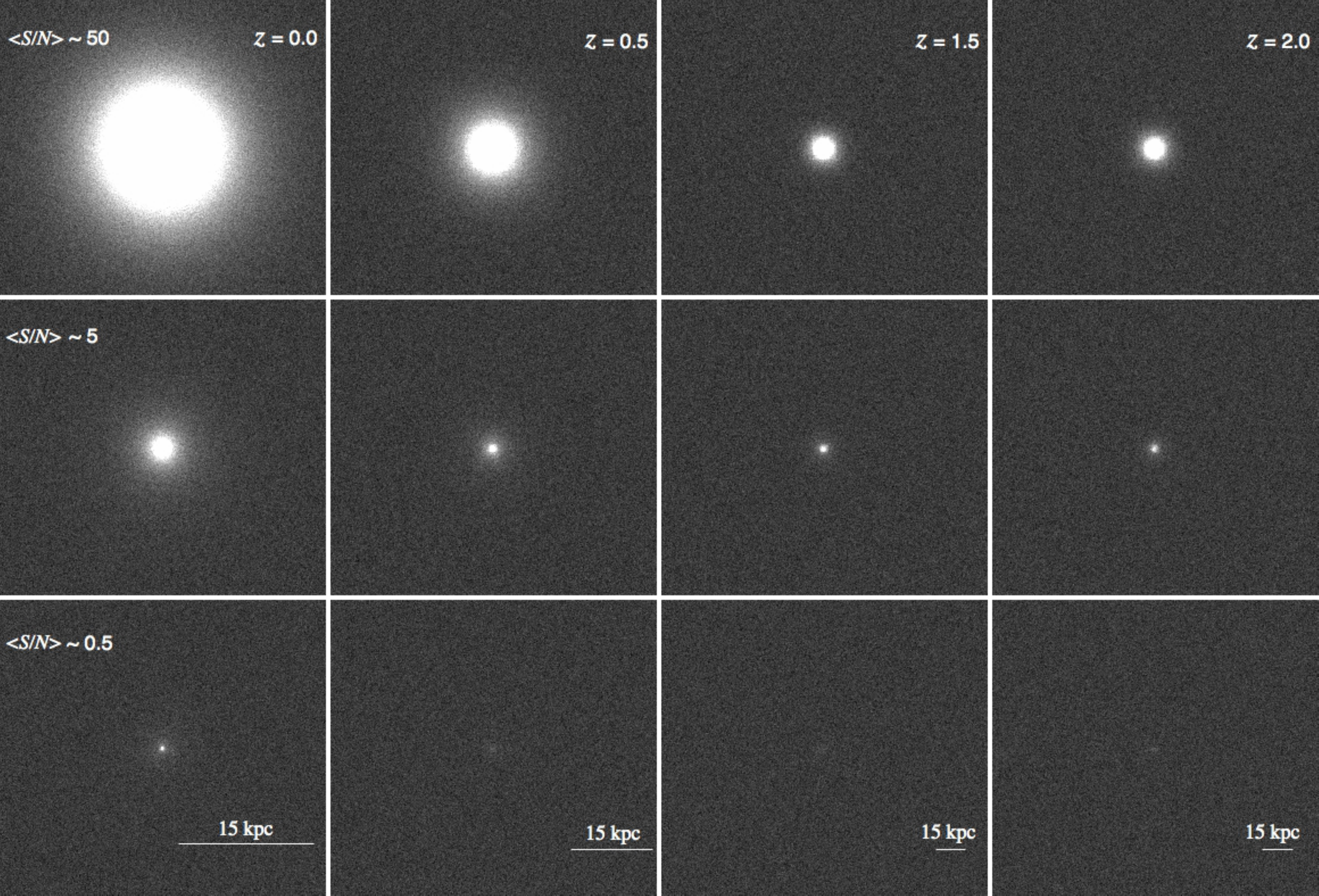}
  \caption{Simulated multi-component galaxies at four different redshifts and three averaged signal-to-noise ratios ($\langle S/N \rangle$).  The model is based on the best-fit model for NGC 1379 from H13. The simulated galaxy at \emph{z} = 2.0 (right columns) has $R_e =$ 6 pixels and $e =$ 0 and therefore this galaxy is shown at $S/N$ $\approx$ 5, 50, and 500.\label{fig:cgs_example}}
\end{figure*}

\begin{equation}
\Sigma(R) = \Sigma_e \ exp {\left \{-\kappa \left [ \left (\frac{R}{R_e} \right )^{1/n} - \  1\right ]\right \}},
\label{eq:sersic}
\end{equation}

\noindent where $R_e$ is the effective radius of the galaxy (equivalent to $r_{50}$, the half-light radius), $\Sigma_e$ is the surface brightness at $R_e$, the S\'{e}rsic index $n$ describes the profile shape, and the parameter $\kappa$ is closely connected to $n$ (\citealt{Ciotti91}). Together with position (\emph{x} and \emph{y}), axis ratio \emph{b/a}, and position angle, this profile has seven free parameters. The S\'{e}rsic profile represents a more general form of the exponential light profile seen in galactic disks (\emph{n} = 1; \citealt{Freeman70}) and the \emph{$R^{1/4}$}-law (de Vaucouleurs law; \citealt{deVaucouleur48}) typical of luminous early-type galaxies (\emph{n} = 4).  Modeling with this profile has been explored in detail in several works (e.g., \citealt{Simard98}; \citealt{Simard02}; \citealt{Graham05}). Many authors have used a constant value of \emph{n} = 2.0 or 2.5 to crudely distinguish early-type (bulge-dominated) from late-type (disk-dominated) galaxies (e.g., \citealt{Blanton03}; \citealt{Shen03}; \citealt{Bell04}; \citealt{Hogg04}; \citealt{Ravindranath04}; \citealt{Barden05}; \citealt{McIntosh05}; \citealt{Fisher08}). S\'{e}rsic profiles with higher S\'{e}rsic indices have longer tails that make the analysis of these galaxies more challenging due to greater sensitivity to neighboring objects, to profile mismatch, and to accurate knowledge of the sky background (see Figure~3 of \citealt{Peng10}). We will compare our results for different ranges of S\'{e}rsic indices. 
 
Sky background and noise are added to all the images in two different ways: (1) adding an artificial background level with Poisson noise using {\tt IRAF}/{\tt mknoise}\footnote{{\tt IRAF} is distributed by the National Optical Astronomy Observatories, which are operated by the Association of Universities for Research in Astronomy, Inc., under cooperative agreement with the National Science Foundation.} (\citealt{Tody86, Tody93}); (2) putting the simulated galaxy on an actually observed background. Although the main focus of this study is galaxies at \emph{z} = 2, our simulations are more generally applicable. The simulations explore a range of fundamental measurables, i.e. sizes, S\'{e}rsic indices, axis ratio, and $S/N$.  As the simulations extend down to instrumental limits, they apply to galaxies at any redshifts provided that one first converts apparent magnitude into $S/N$.  As illustration, we compare galaxies at \emph{z} = 0.5, 1.5, and 2, to our simulations. The simulated observed backgrounds are taken from CANDELS\footnote{\tt http://candels.ucolick.org/} (\citealt{Grogin11}; \citealt{Koekemoer11}) UDS (UKIDSS Ultra-Deep Survey; \citealt{UKIDSS}) $\emph{I}_{814}$, $\emph{J}_{125}$, and $\emph{H}_{160}$ mosaic images, respectively (Figure \ref{fig:skies}). Therefore, when local galaxies are rescaled to specifically match in size and $S/N$ of observed galaxies in those filters, \emph{K}-correction is not required. The simulated backgrounds resemble the observed backgrounds as they have identical gain, pixel scale, and magnitude zero point ({\it magzpt}) as in the UDS mosaic images. Furthermore, the background root-mean-square (RMS; the standard deviation from the median sky value) of the simulated and observed backgrounds are comparable. 

All artificially generated galaxies are convolved with a CANDELS UDS hybrid PSF (\citealt{vanderWel12}) of the filter that corresponds to the redshift of interest. An important factor in obtaining a best-fit model is the accuracy of the PSF. The effective radii of the red nuggets are comparable to and in some cases smaller than the PSF full-width at half maximum (FWHM). Therefore, one may expect some offset in {\tt GALFIT} measurements when an inaccurate PSF is used. This issue is studied in the Appendix, where we show that the effects of using slightly different PSFs are not more than 5$\%$ on the final model.

One of the most important factors in morphology analysis is $S/N$. There are several ways to define $S/N$ because galaxies are extended.  One way, which is somewhat analogous to total $S/N$, is given by 

\begin{equation}
S/N = \frac{f_{\rm galaxy}}{\sqrt{f_{\rm galaxy}+A\sigma^2}},
\label{eq:sn}
\end{equation}

\noindent
where three different parameters determine the $S/N$: the area $A$ of the aperture in which the $S/N$ is measured, the galaxy total flux $f_{\rm galaxy}$ within $A$, and the background RMS $\sigma$. The RMS here is the sum of all possible sources of noise, including shot noise from the sky, readout noise, and shot noise in dark current. 
In this paper, $A$ includes every pixel within $R_e$. Note that this is different from measuring $S/N$ within an aperture with a constant size because the aperture size here is different for each galaxy. Equation \ref{eq:sn}, however, may not be the most informative. For various technical reasons, it is harder to measure structural properties of low-surface brightness galaxies. Thus, another useful definition is analogous to an average surface brightness, defined as $\langle S/N \rangle$ = $\frac{S/N}{\sqrt{\pi{R_e}^{2}(1-e)}}$.  Here, the ellipticity $e = 1-b/a$.  Figure 2 shows a set of galaxies and the corresponding $S/N$ definitions.

By adding galaxies to an observed background, one can study the effect of faint undetected objects or being in the vicinity of bright and/or extended objects. One way to minimize the effects of neighboring objects is to mask them out.  We accomplish this by running SExtractor\footnote{\tt http://www.astromatic.net/software/sextractor} (\citealt{Bertin96}) with the parameter \emph{Detection$\_$threshold} set to 1.5. Instead of using the SExtractor segmentation map, the location ({\tt XPEAK\_IMAGE, YPEAK\_IMAGE}), ellipticity, position angle, and the area of the objects from the SExtractor output catalog are used to create a bad pixel mask image. First, the semi-major axis of each galaxy is calculated by using the measured area and ellipticity of the object. Bad pixel regions are constructed by assigning the center, the semi-major axis, the semi-minor axis, and the position angle of each object. We double the calculated semi-major and semi-minor axes (i.e., quadruple the area) in order to mask out most of the faint outer parts of the objects. Figure \ref{fig:skies} shows the observed backgrounds and bad pixel masks. These masks are used for {\tt GALFIT} modeling. While masking does not fully eliminate the possibility of neighboring contamination (see \citealt{Haussler07}), this method is analogous to that of some studies in the literature and is thus informative.  Together with analysis on idealized backgrounds, these two scenarios give an idea of the different levels of systematics.

\subsection{Single-component Galaxies}

Simulating single-component galaxies is one of the most basic, yet important ways of testing the robustness of fitting pipelines (\citealt{Haussler07}). We simulate more than 10,000 single-component galaxies. The S\'{e}rsic component parameters are randomly chosen from the following ranges: 0.8 $<$ $R_e$ $<$ 35 pixels ($\sim$0\farcs05--2\farcs0 in WFC3/$H_{160}$), 0.5 $<$ $n$ $<$ 5, and 0 $<$ $e$ $<$ 0.8. For one {\it HST}\ orbit,  a $S/N$ range between 1 and 1000 within the effective radius of the simulated galaxies roughly corresponds to a magnitude range of 20 $<$ $m_H$ $<$ 27.

Once the galaxies are simulated, we fit each with a single S\'{e}rsic component plus a sky component. For the galaxy component, the initial parameter guesses include the position of the galaxy (physical $x$ and $y$), S\'{e}rsic index, magnitude, effective radius, ellipticity, and position angle.

\subsection{Multi-component Galaxies}

Real galaxies have structures that complicate the analysis.  Several recent studies looked into this issue by creating bulge + disk models, and then fitting single-component S\'{e}rsic models to them (\citealt{Meert13}; \citealt{Mosleh13}).  The question was whether the disk components were missed by the analysis technique, therefore leading to underestimation of the sizes.  They found that single-component models were able to recover the magnitude and sizes, with only small systematic errors (0\%--20\%), and that the multi-component analysis performed better.

While those studies are informative in their own right, there is another complementary approach.  Our simulations are designed to more directly test the null hypothesis that elliptical galaxies today are the direct descendants of \emph{z} $\approx$ 2 galaxies that have undergone no morphological changes in structure or size.  To do so, we start with 100 nearby elliptical galaxies with multi-component decomposition models from H13 that capture the detailed structures down to very low surface brightnesses, accounting for the PSF.   We then rescaled the composite (sum of the multiple components) as a single unit to span a range of sizes and signal-to-noise similar to galaxies at \emph{z} $\approx$ 2.  We convolve the composite model with a PSF and then perform single-component S\'{e}rsic fits to it. The rationale for this approach is to be agnostic about the structural nature of high-\emph{z} systems (e.g., whether they are disks or bulges) because that judgment may itself be entangled with the technique as well as other more subtle technical nuances.

We rescale the sizes and luminosities of the H13 multi-component models to generate about 300 multi-component galaxies with $R_e \approx 1-70$ pixels and $S/N$ from a few to 1000. This range is relevant for studying massive early type galaxies down to the resolution and $S/N$ limits of {\it HST}\ observations, regardless of redshift. Images of galaxies analogous to being at \emph{z} = 0.5, 1.5, and 2.0 are simulated based on the $S/N$ properties of $\emph{I}_{814}$, $\emph{J}_{125}$ , and $\emph{H}_{160}$ (equivalent to the rest-frame optical) CANDELS UDS images, respectively.

As in the single-component simulations, we add either a simulated background or an observed background. By using 300 multi-component galaxies as templates, more than 6000 galaxies with 5 $<$ $S/N$ $<$ 1000 (uniformly distributed in logarithmic space) are simulated.

We analyze the multi-component galaxies in a manner similar to that used to treat the single-component galaxies. To replicate the most commonly adopted analysis method, we fit every galaxy with a single S\'{e}rsic and sky component.  However, because of the mixture of different components, another method is needed to determine the intrinsic size and other parameters of the galaxy. To establish the baseline for comparison of sizes and other parameters, we employ {\tt IRAF}/{\tt ellipse} (\citealt{Jedrzejewski87}), a task that fits elliptical isophotes to images, to create a curve of growth (CoG) of the cumulative radial flux distribution of the model galaxy.  This is generated with an ellipticity and position angle fixed to the radially average value. No noise is added, and PSF convolution is not applied.  As the total flux of the galaxy is known, the CoG allows us to obtain a non-parametric estimate of the effective radius, the radius at which half of the total flux is enclosed.
 
\begin{figure}[t]
  \centering
\includegraphics[width=95mm]{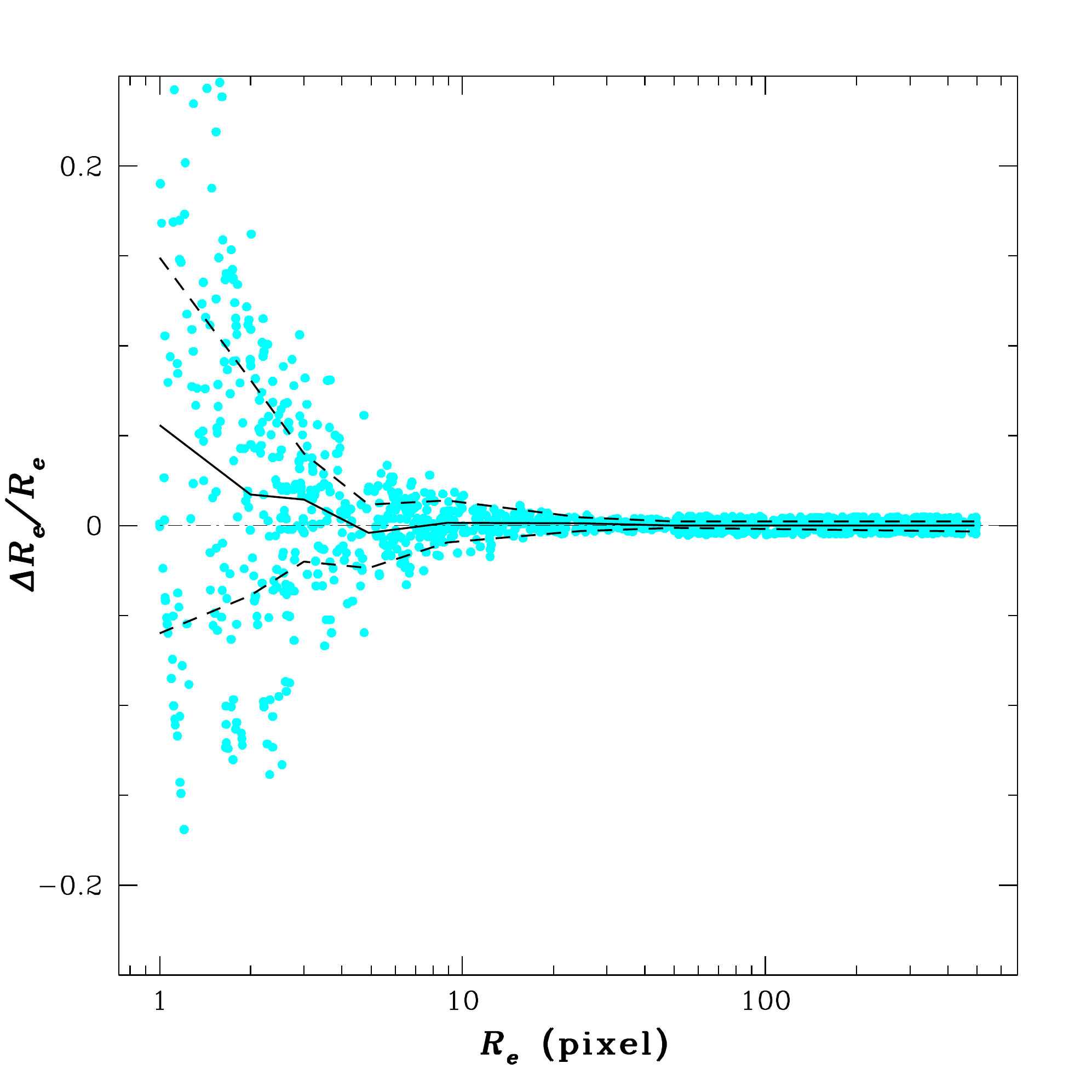}
  \caption{Testing the reliability of size measurements using curve of growth (CoG) analysis. The $x$-axis shows the effective radius of 1250 single-component galaxies, and the $y$-axis shows the difference between the actual size and the size measured by CoG. Black solid and dashed lines indicate the median and 1$\sigma$ uncertainties. The dot-dash line shows zero offset. The size of galaxies with 1 $<$ $R_e$ $<$ 2 pixels can be overestimated by 5\% and with 10\% uncertainties. For galaxies with 2 $<$ $R_e$ $<$ 5 pixels, the sizes can be measured without any bias and with 5\% uncertainties. And for larger galaxies the uncertainties are less than 1\% with no bias.\label{fig:ellipse}}
\end{figure}

\begin{figure*}
  \centering
\includegraphics[width=139mm]{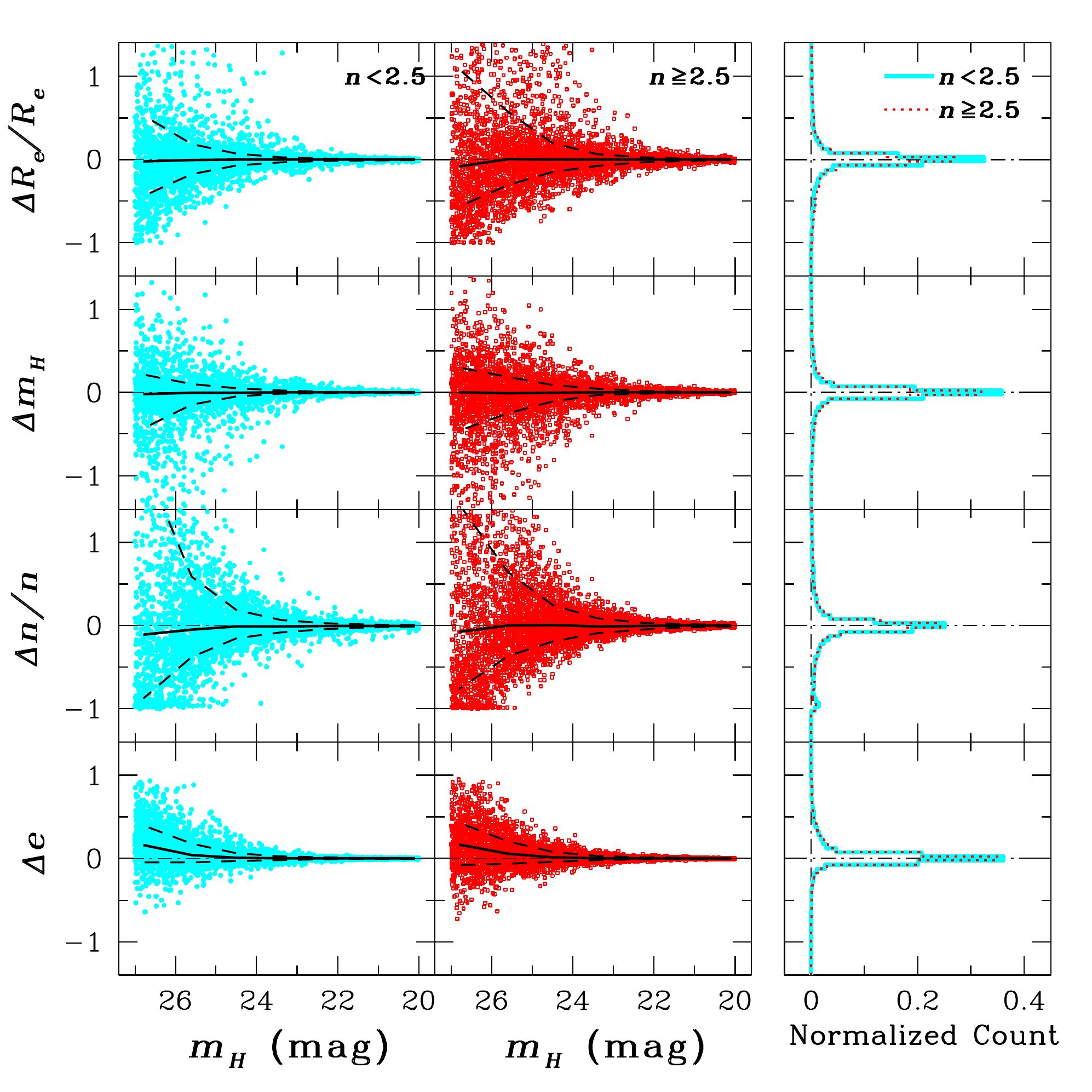}
  \caption{The results of more than 10,000 single-component model galaxies with a simulated background. From top to bottom, panels show the offsets between the measured and the actual effective radius $R_e$, magnitude $m_H$ (in one {\it HST}\ orbit), S\'{e}rsic index $n$, and ellipticity $e$, respectively. Black solid and dashed lines in the scatter plots indicate the median and 1$\sigma$ uncertainties of different measurements. The dot-dash lines show the zero offset. Cyan points and cyan solid lines show the results of galaxies with input $n$ $<$ 2.5, and red points and red dotted lines are for galaxies with $n$ $\geq$ 2.5. The median offsets are zero down to $m_H \approx 26$, but the scatter rises steeply in the magnitude range between 25 and 26. \label{fig:1comp1}}
\end{figure*}

\begin{figure*}
  \centering
\includegraphics[width=126mm]{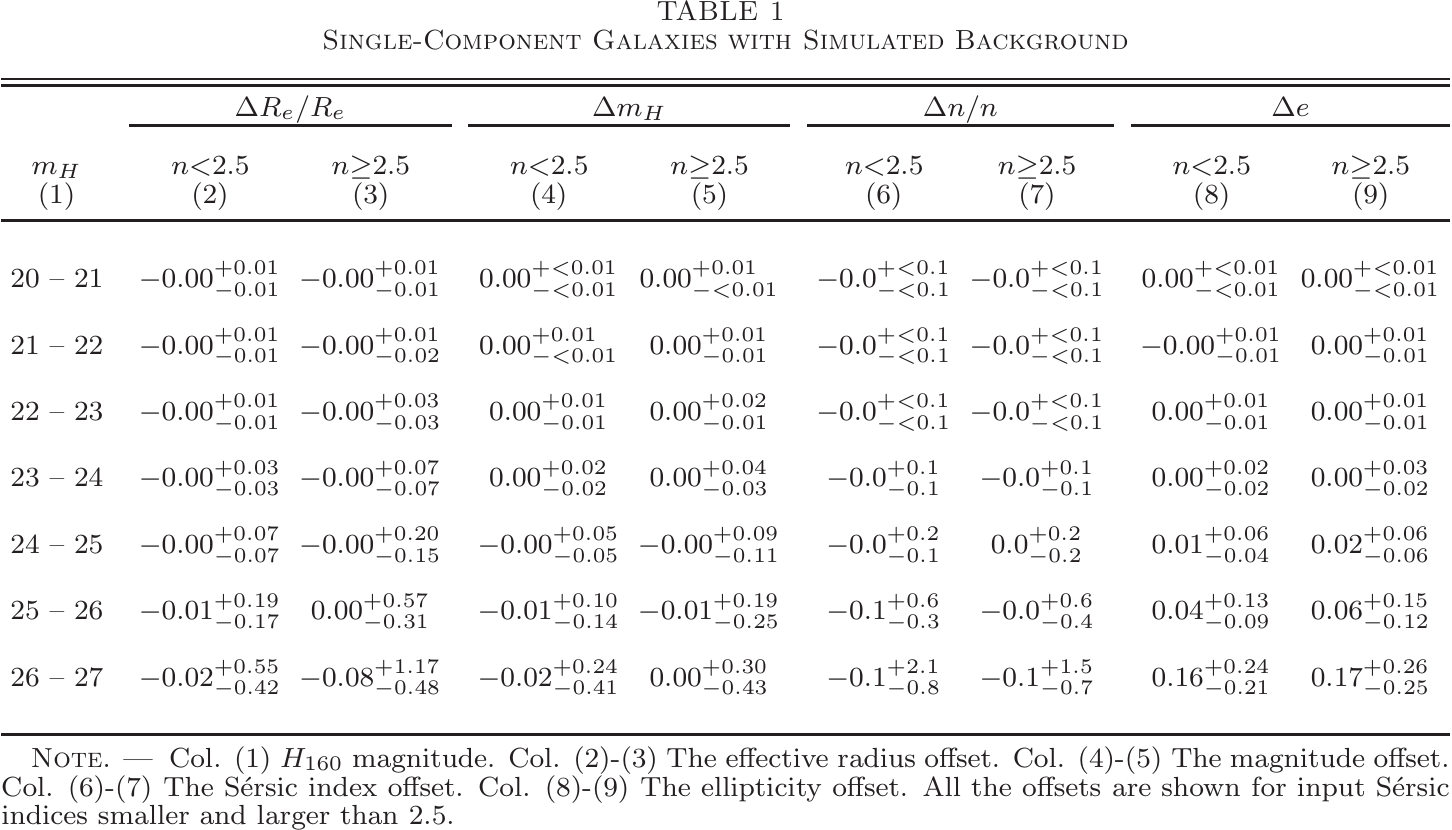}
\end{figure*}

\begin{figure*}
  \centering
\includegraphics[width=139mm]{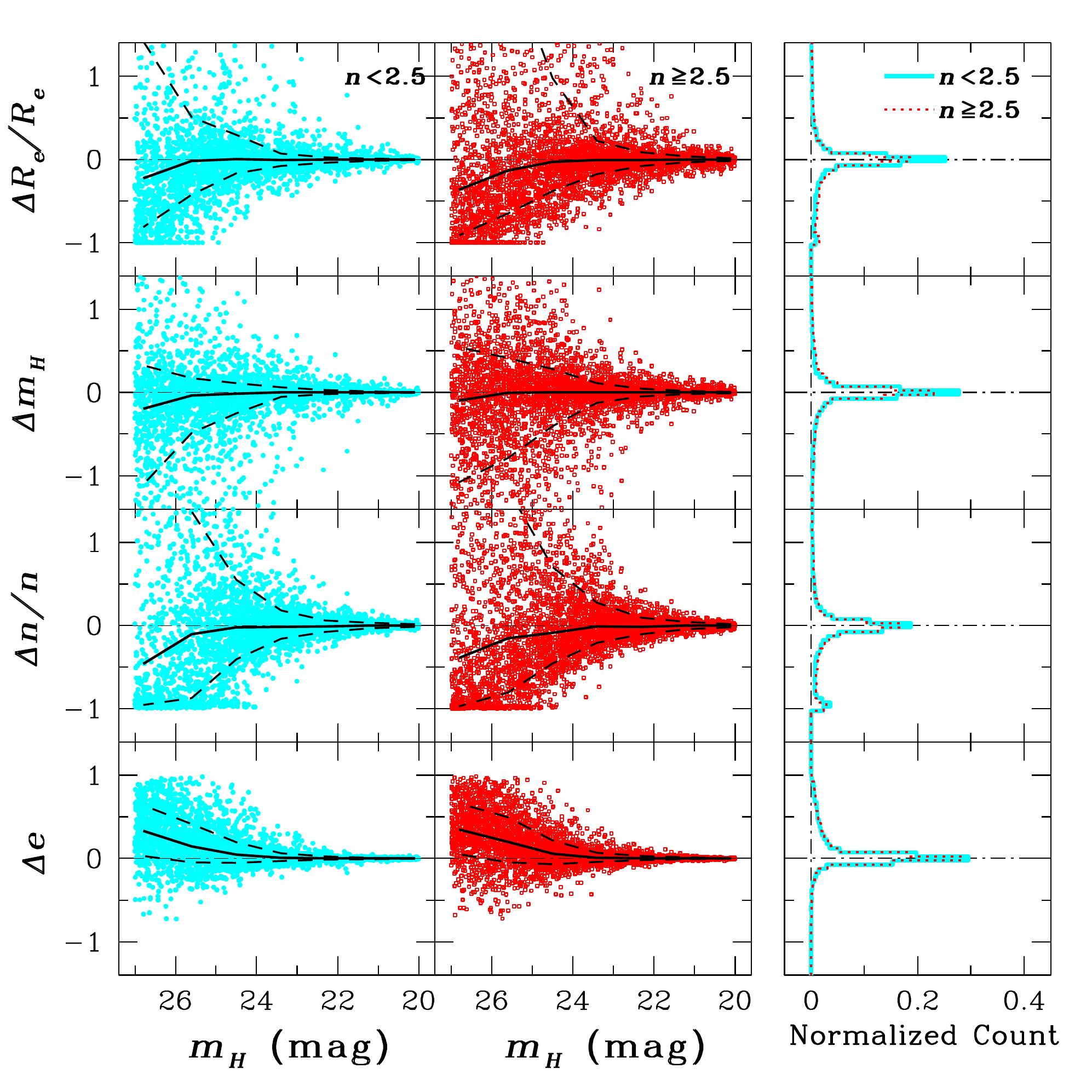}
  \caption{The results of more than 10,000 single-component model galaxies with an observed background. From top to bottom, panels show the offsets between the measured and the actual effective radius $R_e$, magnitude $m_H$ (in one {\it HST}\ orbit), S\'{e}rsic index $n$, and ellipticity $e$, respectively.  Black solid and dashed lines in the scatter plots indicate the median and 1$\sigma$ uncertainties of different measurements. The dot-dash lines show the zero offset. Cyan points and cyan solid lines show the results of galaxies with input $n$ $<$ 2.5, and red points and red dotted lines are for galaxies with $n$ $\geq$ 2.5. The median offsets are zero down to $m_H \approx 25$, but the scatter rises steeply in the magnitude range between 24 and 25. \label{fig:1comp2}}
\end{figure*}

\begin{figure*}
  \centering
\includegraphics[width=126mm]{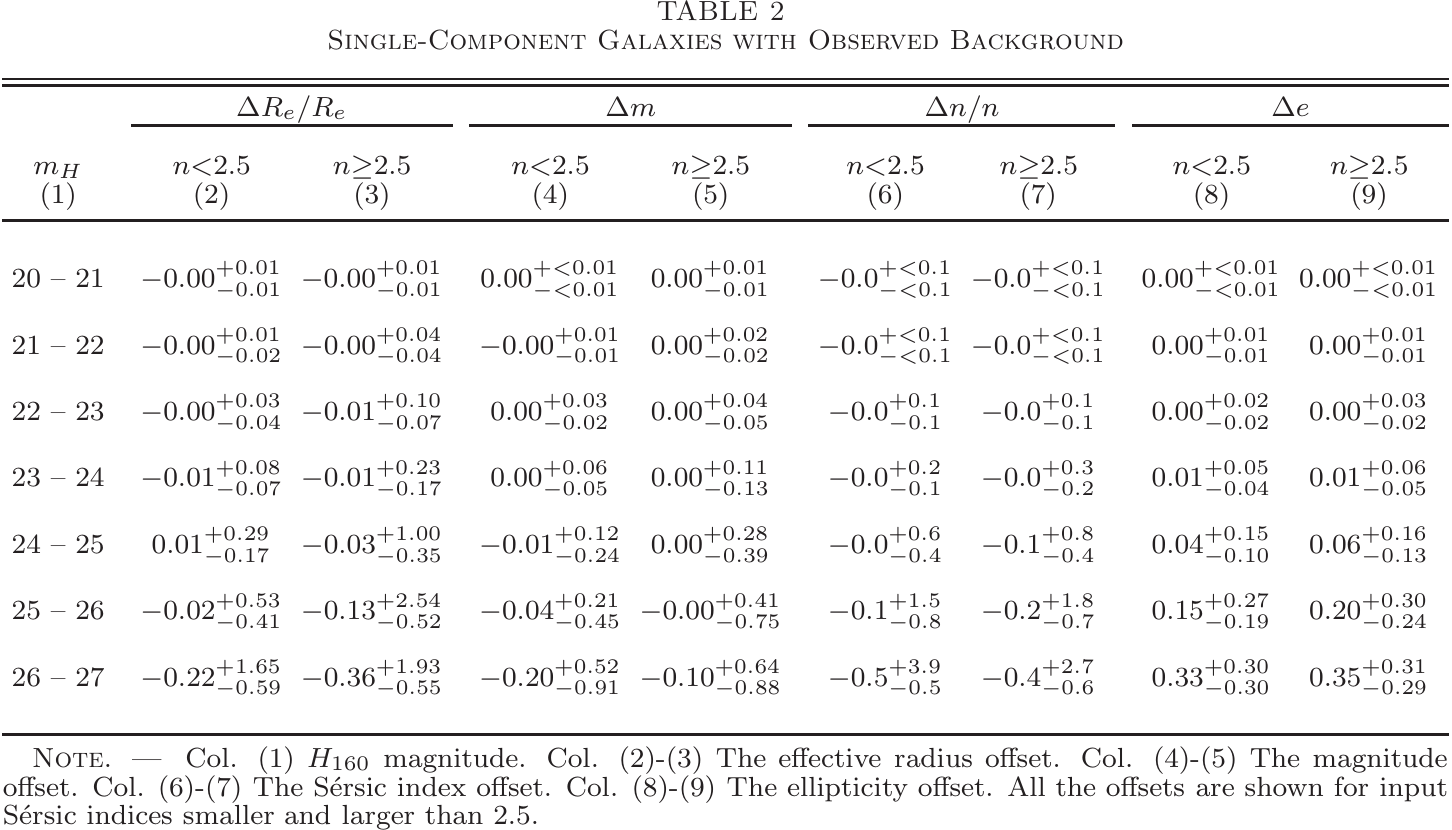}
\end{figure*}

\begin{figure*}
  \centering
\includegraphics[width=137mm]{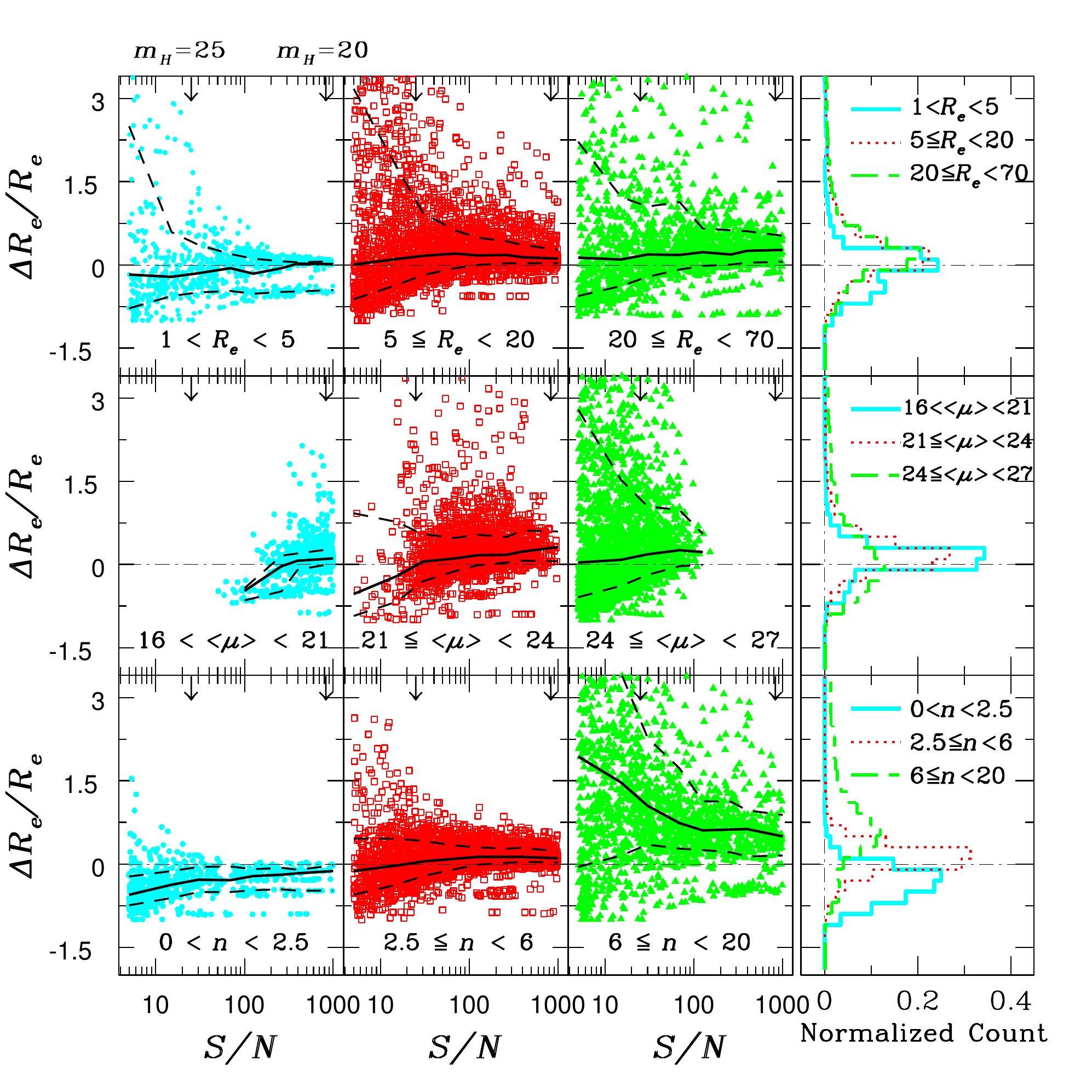}
  \caption{Results of more than 7,000 multi-component galaxies with simulated backgrounds. The upper, middle, and lower panels show the results at different intervals of size $R_e$ (pixel), mean surface brightness $\langle \mu \rangle$ ($m_H$  arcsec$^{-2}$), and S\'{e}rsic index $n$, respectively.  Black solid and dashed lines in the scatter plots indicate the median and 1$\sigma$ uncertainties of different measurements. The dot-dash lines show the zero offset. The downward-pointing arrows in the scatter plots indicate the $S/N$ of a galaxy with $R_e$ = 5 pixels, $e$ = 0, and $m_H$ = 20 and 25. Note the significant size and total luminosity overestimations of {\tt GALFIT} models with $n$ $>$ 6; see text in Section 3.2 regarding the offsets in different bin internals of S\'{e}rsic index. \label{fig:3comp1}} 
\end{figure*}

\begin{figure*}
  \centering
\includegraphics[width=130mm]{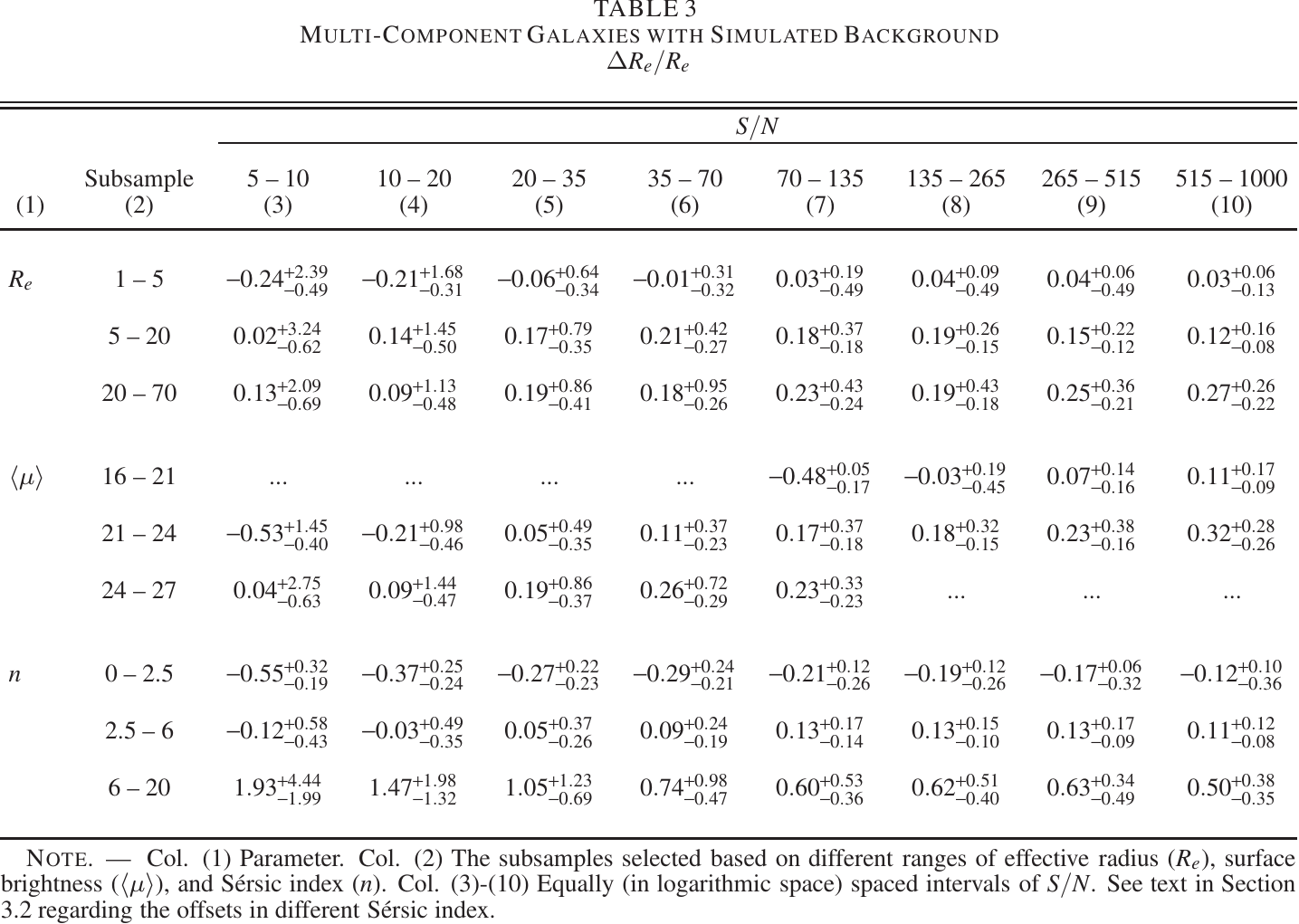}
\end{figure*}

\begin{figure*}
  \centering
\includegraphics[width=130mm]{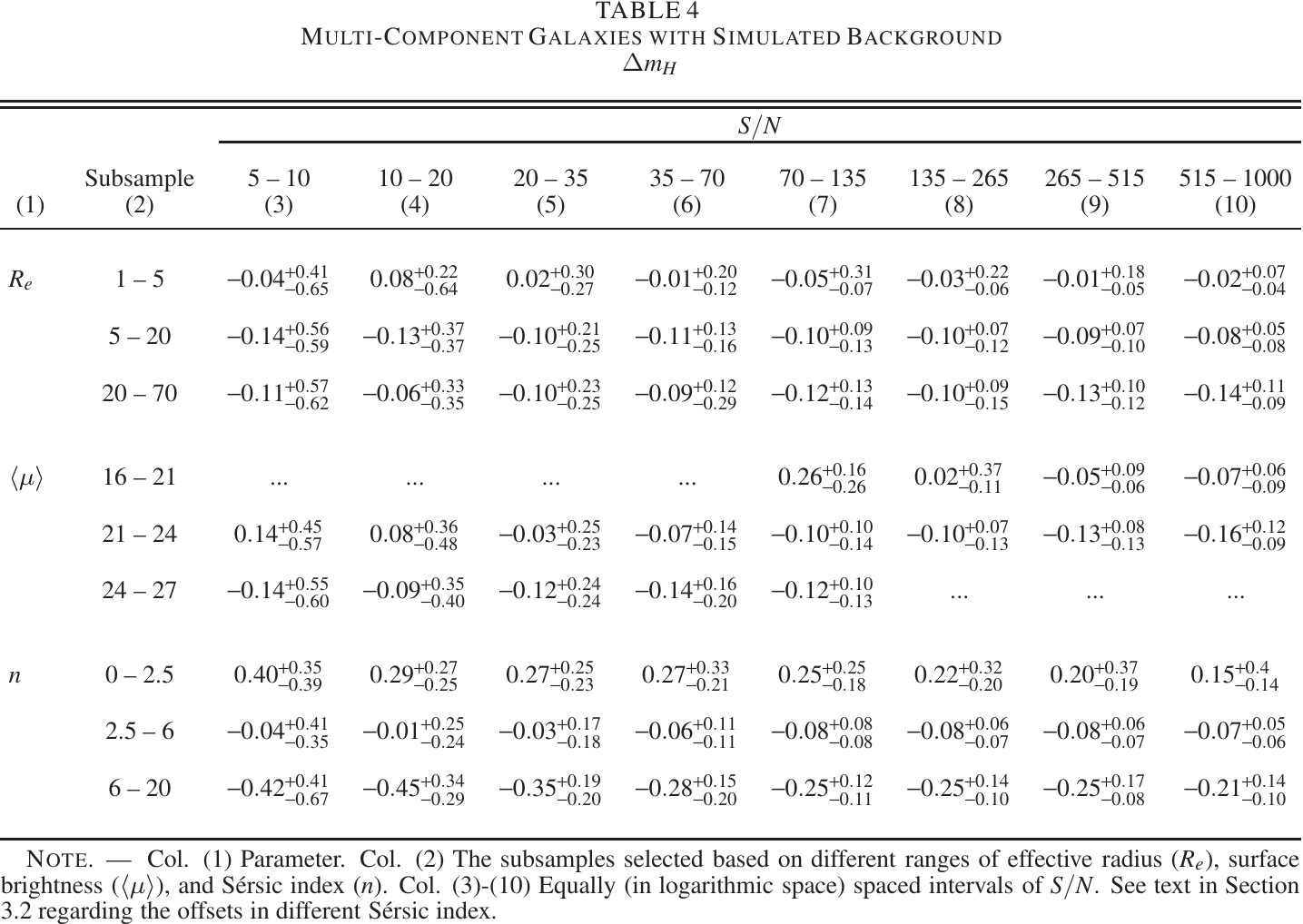}
\end{figure*}

\begin{figure*}
  \centering
\includegraphics[width=137mm]{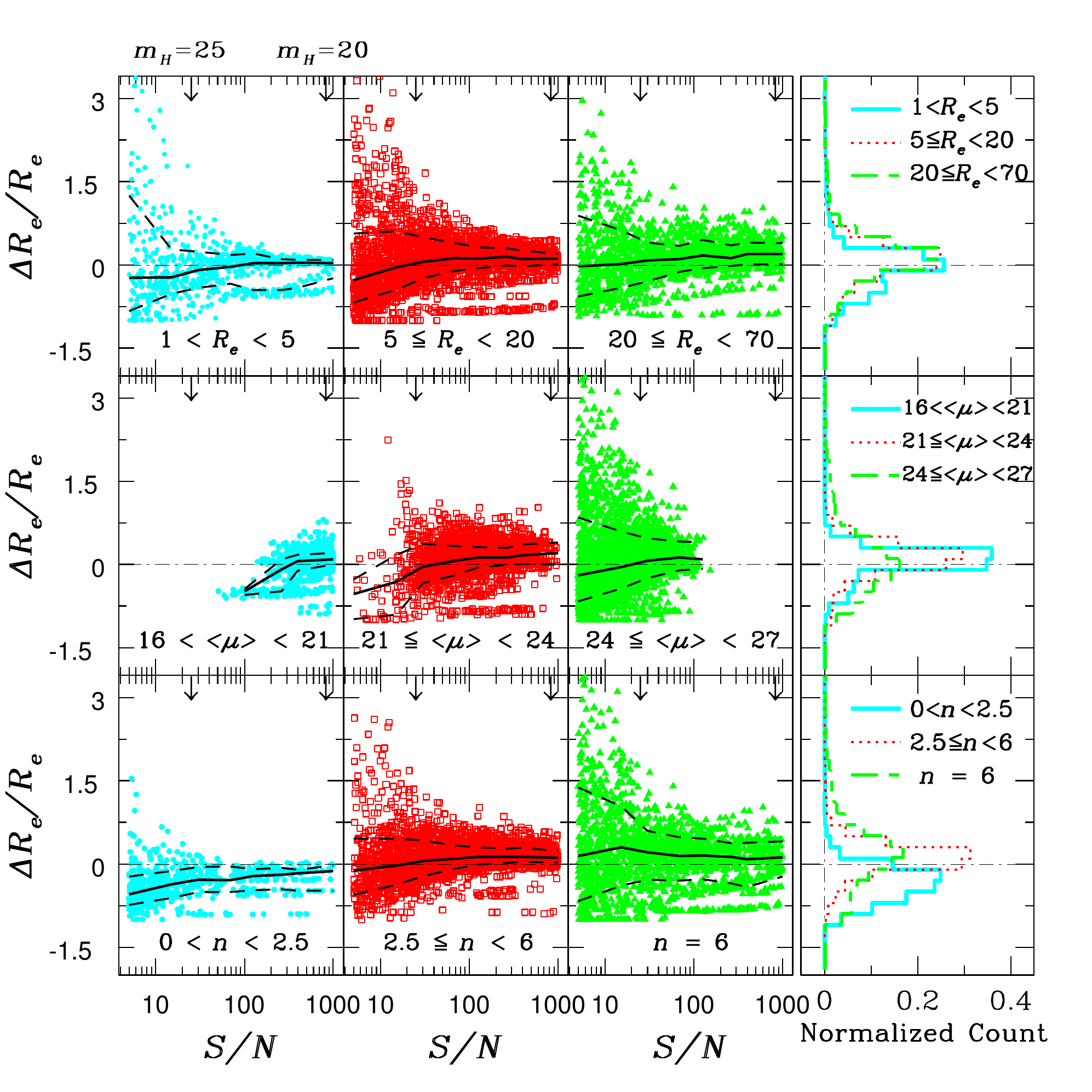}
  \caption{Results of more than 7,000 multi-component galaxies with a simulated background. The upper, middle, and lower panels show the results at different intervals of size $R_e$ (pixel), mean surface brightness $\langle \mu \rangle$ ($m_H$  arcsec$^{-2}$), and S\'{e}rsic index $n$, respectively.  Black solid and dashed lines in the scatter plots indicate the median and 1$\sigma$ uncertainties of different measurements. The dot-dash lines show the zero offset. The downward-pointing arrows in the scatter plots indicate the $S/N$ of a galaxy with $R_e$ = 5 pixels, $e$ = 0,  and $m_H$ = 20 and 25. Galaxies with $n$ $>$ 6 are refit by fixing the S\'{e}rsic index to $n$ = 6. Note the significant decrease in scatter and systematic errors (compare with Figure \ref{fig:3comp1}); see text in Section 3.2 regarding the offsets in different bin internals of S\'{e}rsic index. \label{fig:3comp2}}
\end{figure*}

\begin{figure*}
  \centering
\includegraphics[width=130mm]{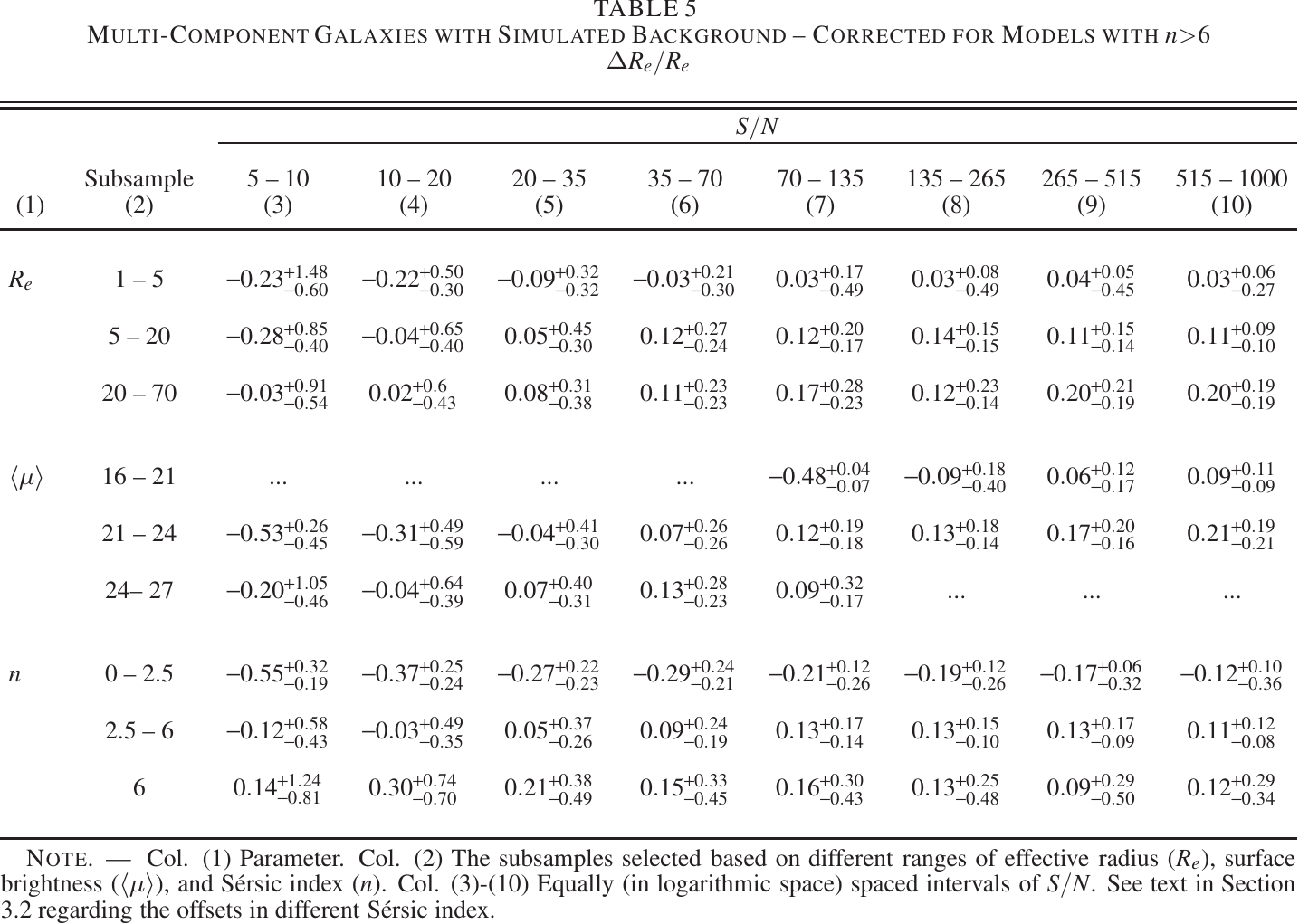}
\end{figure*}

\begin{figure*}
  \centering
\includegraphics[width=130mm]{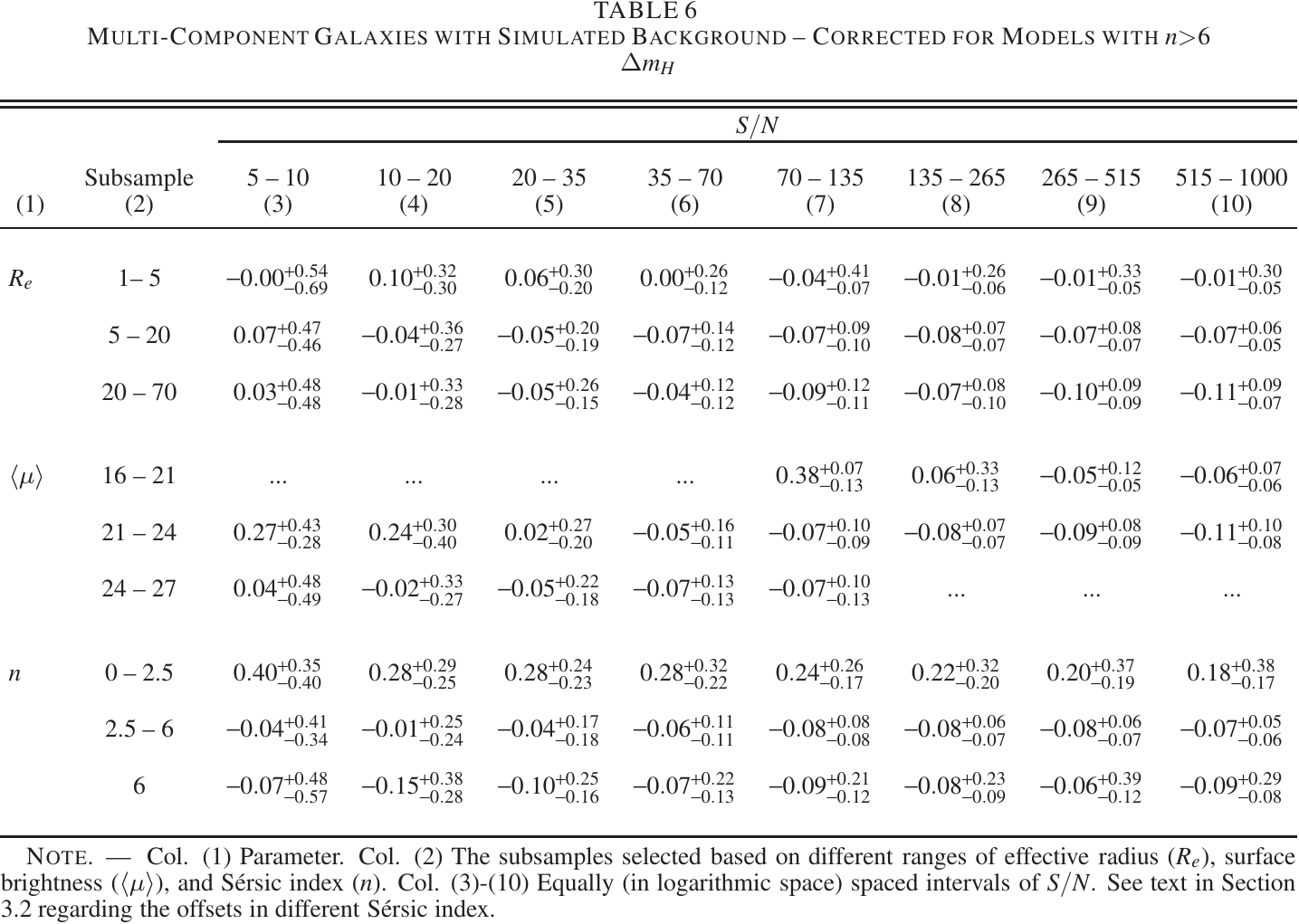}
\end{figure*}

\begin{figure*}
  \centering
\includegraphics[width=137mm]{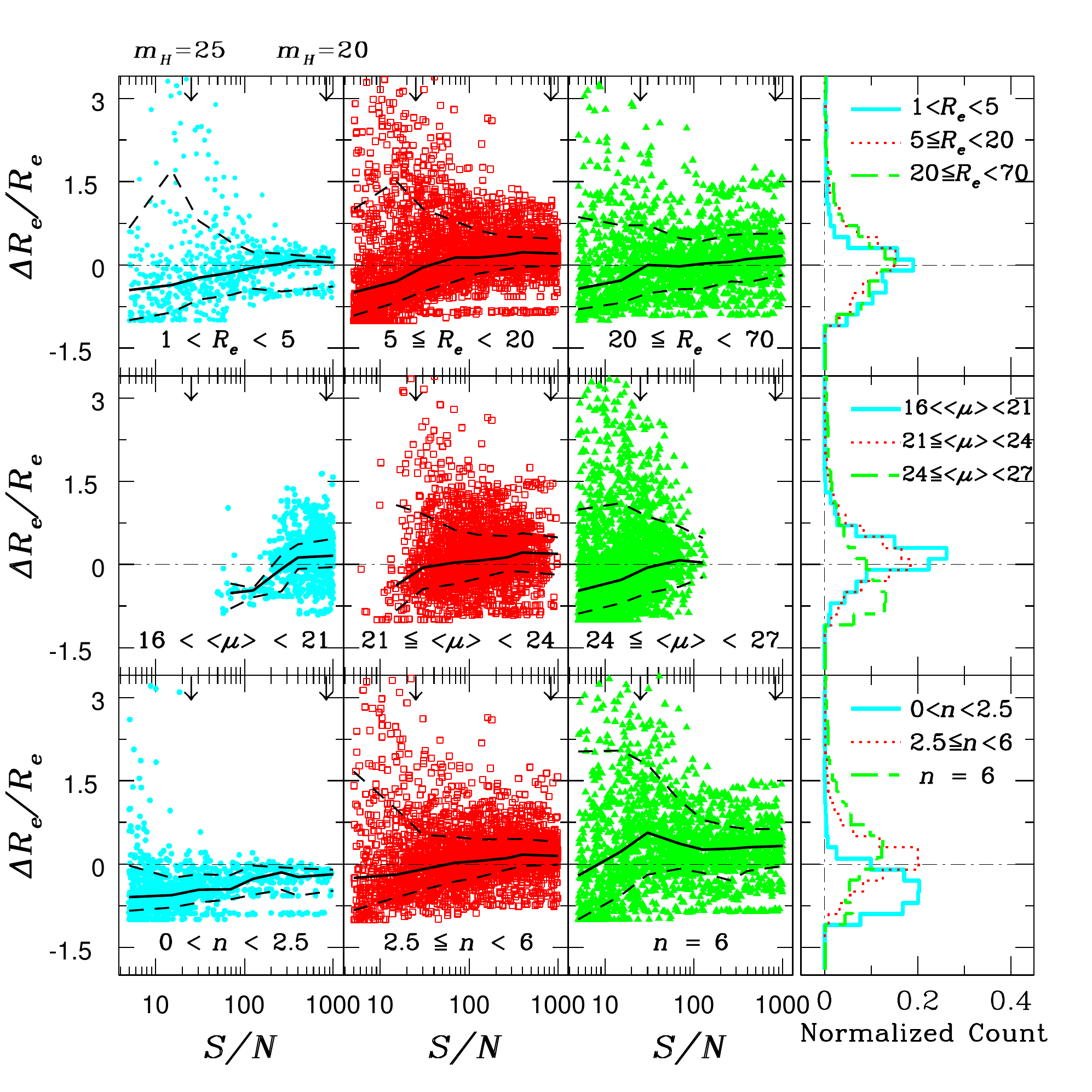}
  \caption{Results of more than 7,000 multi-component galaxies with an observed background. The upper, middle, and lower panels show the results at different intervals of size $R_e$ (pixel), mean surface brightness $\langle \mu \rangle$ ($m_H$  arcsec$^{-2}$), and S\'{e}rsic index $n$, respectively.  Black solid and dashed lines in the scatter plots indicate the median and 1$\sigma$ uncertainties of different measurements. The dot-dash lines show zero offset. The downward-pointing arrows in the scatter plots indicate the $S/N$ of a galaxy with $R_e$ = 5 pixels, $e$ = 0,  and $m_H$ = 20 and 25. Galaxies with $n$ $>$ 6 are refit by fixing the S\'{e}rsic index to $n$ = 6. The trends are similar to the results for simulated backgrounds, although the scatter is larger; see text in Section 3.2 regarding the offsets in different bin intervals of S\'{e}rsic index. \label{fig:3comp3}}
\end{figure*}

\begin{figure*}
  \centering
\includegraphics[width=130mm]{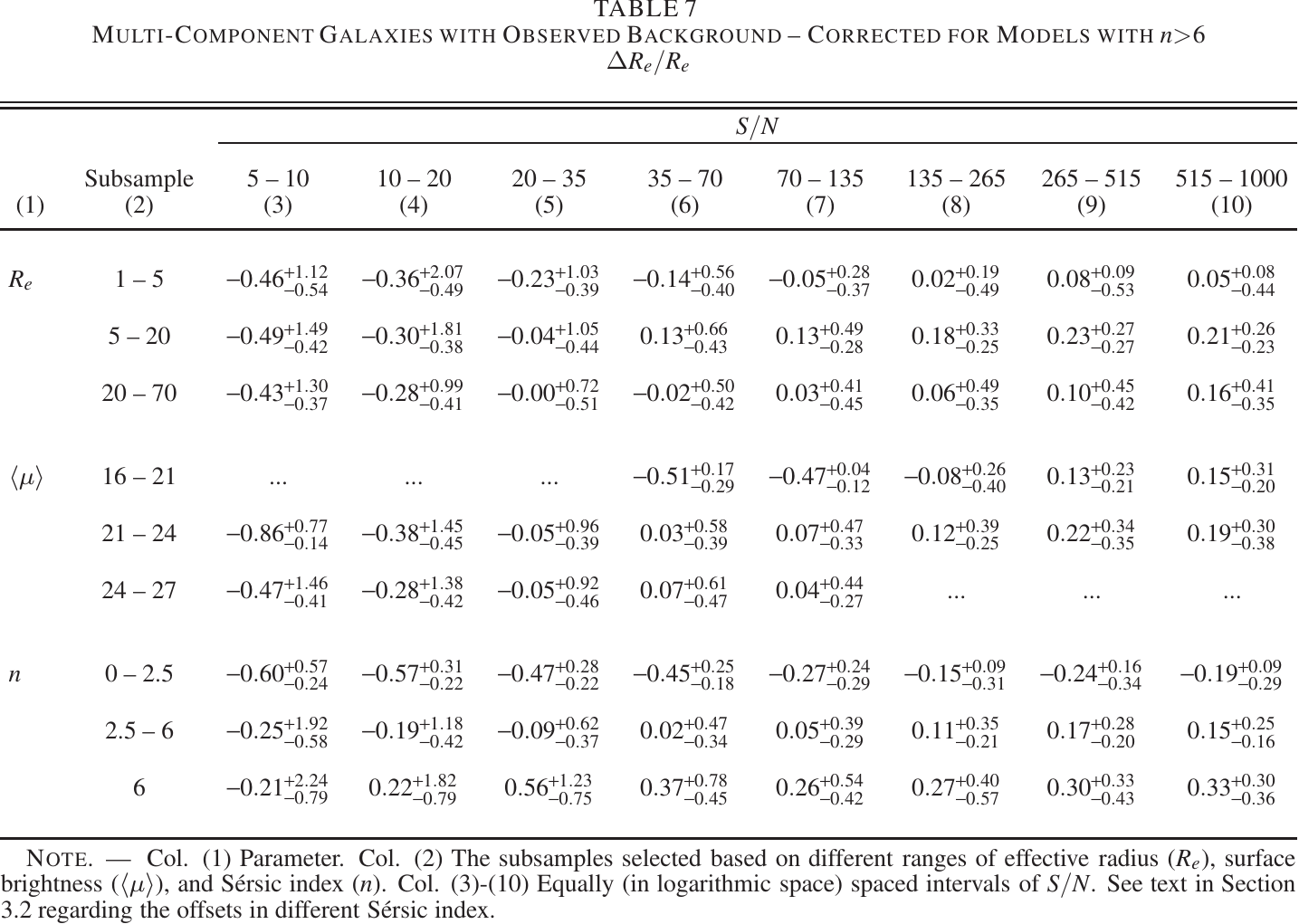}
\end{figure*}

\begin{figure*}
  \centering
\includegraphics[width=130mm]{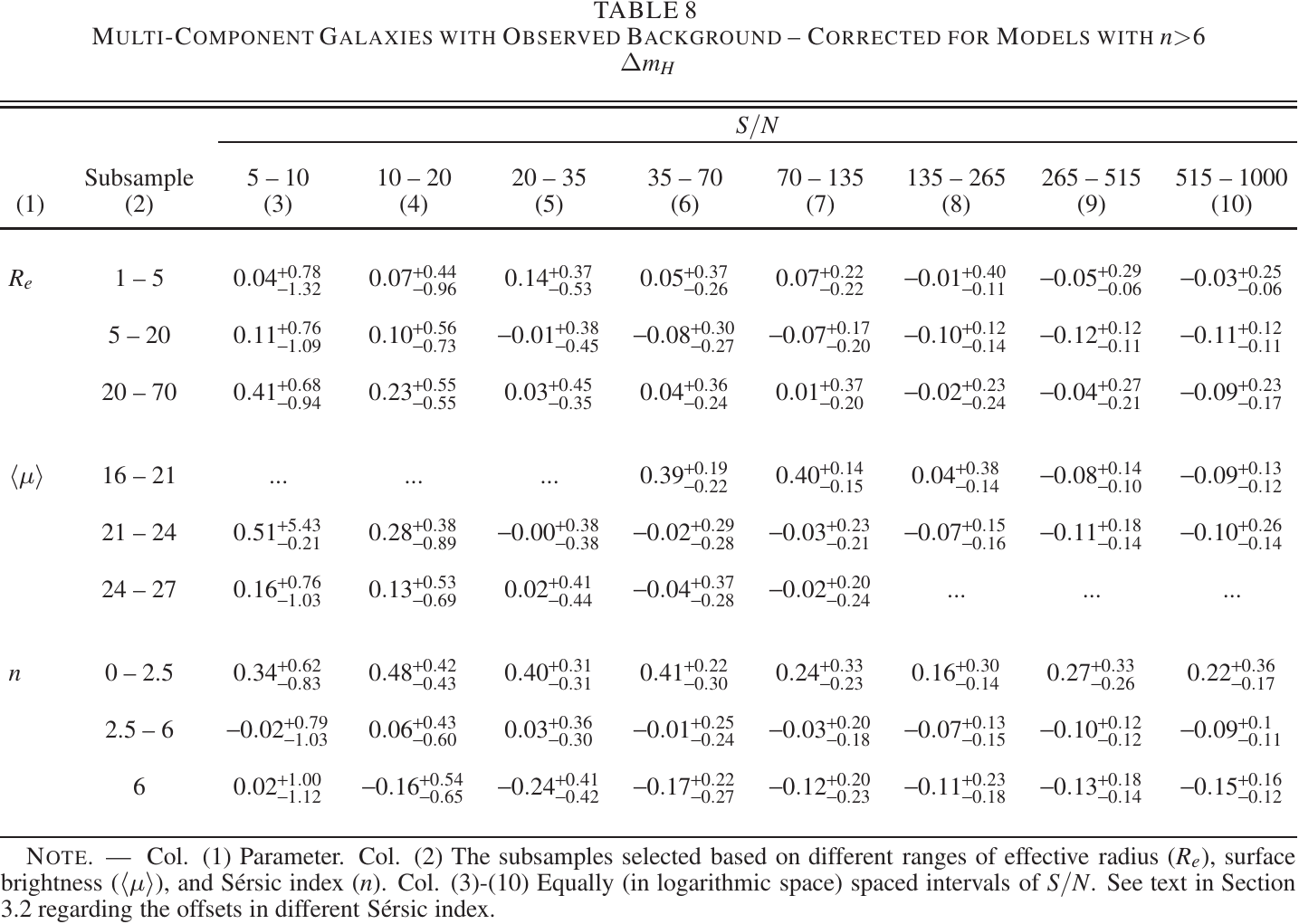}
\end{figure*}

To examine the robustness of our CoG analysis, we employ this method to measure the sizes of 1250 single-component galaxies with a large range of sizes (1--500 pixels). Figure \ref{fig:ellipse} shows that there is no bias in size measurements for the larger galaxies but the finite size of the pixels makes size determination of the smallest galaxies slightly uncertain. 
Sizes with 1 $<$ $R_e$ $<$ 2 pixels can be overestimated by 5$\%$ and with 10$\%$ uncertainty. For galaxies with 2 $<$ $R_e$ $<$ 5 pixels, the sizes can be measured without any bias and with 5\% uncertainty. And for larger galaxies the uncertainties are less than 1\% with no bias. We take these effects into consideration when analyzing the multi-component galaxies. However, fewer than 2\% of the simulated galaxies have $R_e$ $<$ 3 pixels. 

Lastly, we note that because the CoG technique averages over all elliptical annuli to measure the effective radius, the comparison between CoG and single-component S\'{e}rsic fits in principle is imperfect for multi-component galaxies whose isophotes often change in ellipticity with radius.  Nevertheless, the general agreement  in the size measurements between the two methods is reassuring.

\section{Results}

\subsection{Single-component Galaxies}

Figure \ref{fig:1comp1} and Table 1 summarize the results of single-component galaxies with a simulated background; these are the most idealized simulations to establish the baseline behavior of the analysis technique for the $S/N$ range of interest. Black solid and dashed lines in the scatter plots indicate the median and 1$\sigma$ uncertainties of different measurements. Cyan points and cyan solid lines show the results of galaxies with input $n$ $<$ 2.5, and red points and red dotted lines are for galaxies with $n$ $\geq$ 2.5. 

To be consistent with the multi-component models later on, where effective radius is more ambiguous to define, the analyzed effective radii are circularized effective radii, $R_{e,{\rm circularized}} \equiv R_{e,{\rm GALFIT}} \times \sqrt{1-e}$.

We see no systematic errors in size, total luminosity, S\'{e}rsic index, or ellipticity measurements down to $m_H$ = 26. And even at 26 $<$ $m_H$ $<$ 27 the systematic errors are less than $\sim$10$\%$. As expected, the uncertainties increase rapidly with decreasing $S/N$ and particularly for $m_H$ $>$ 26.  The S\'{e}rsic index is most vulnerable to large uncertainties.  The simulated images resemble single-orbit {\it HST}/WFC3 $H_{160}$ observations; a galaxy with $R_e$ = 5 pixels (i.e., 0\farcs3 or 2.5 kpc at \emph{z} = 2), $e$ = 0, and $m_H$ = 26 has  $S/N \approx 15$ within $R_e$. 

The uncertainties are, on average, higher for galaxies with larger S\'{e}rsic indices but the systematic offsets are comparable at different $n$ (note that all the simulated single-component galaxies have $n$ $<$ 5). Most of the galaxies with the largest scatter within a magnitude range are ones with the largest sizes, as they have lower mean surface brightness compared to the smaller galaxies.  

Figure \ref{fig:1comp2} and Table 2 summarize the results of  single-component galaxies with an observed background, derived from real images. The scatter is larger but systematic errors are still absent down to  $m_H \approx 25$. One of the main sources of the increased scatter is the presence of bright and/or extended objects in the actual images of the observed background (Figure \ref{fig:skies}). 

It is worth pointing out that masking out objects (using a bad pixel mask for {\tt GALFIT} fitting) is not the most effective way to mitigate the effects of neighboring objects. Instead, one needs to fit the target galaxy and its neighboring objects simultaneously (e.g., \citealt{Haussler07}; \citealt{Barden12}). The properties (e.g., magnitude, size, and S\'{e}rsic index) of the neighboring objects can affect the fit of the target object. In particular, neighbors with the largest $n$ $\times$ $R_e$ have the largest effects.   

Another important factor to consider is that we do not provide an input sigma image (i.e., noise map) for {\tt GALFIT}; instead, we allow {\tt GALFIT} to calculate it.  CANDELS images are generated through an extensive process from raw, single exposures to final, drizzled mosaics. The noise tends to be correlated after the {\tt multidrizzle} procedure, which may lead to underestimating the noise in actual data (\citealt{MultiDrizzle}).  The sigma image generated by {\tt GALFIT} may not include all the necessary information about the image characteristics required for a faithful noise map. \citet{vanderWel12} have analyzed the CANDELS UDS $H_{160}$ image and note that the total flux of objects with $m_H \approx 22 - 23$ and the size of typical objects in their sample (i.e., $\sim$0\farcs3 or $\sim$5 pixels) correspond to the typical background flux level measured within the effective radius. Although our images with simulated backgrounds have similar RMS to CANDELS UDS $H_{160}$ image, the galaxy flux and the background flux are comparable at $m_H \approx 23.5$ for images with simulated backgrounds. This can explain some of the discrepancies between Figure \ref{fig:1comp1} and Figure \ref{fig:1comp2}, even at the bright end.

\subsection{Multi-component Galaxies}

Simulated galaxies with multiple components, where the subcomponents are taken from accurate decompositions of nearby galaxies of H13, provide a better description of the observed galaxy structures than idealized single-component models.  We examine the reliability of measuring the size and total luminosity of multi-component galaxies by single-component fitting.

Figure \ref{fig:3comp1} and Tables 3 and 4 show the difference between actual and measured effective radii at different $S/N$ ranges for about 7,000 model galaxies in images with simulated (i.e. idealized) backgrounds. The upper, middle, and lower panels give the results at different intervals of size, mean surface brightness, and S\'{e}rsic index, respectively.  Black solid and dashed lines in the scatter plots indicate the median and 1$\sigma$ uncertainties of different measurements. The downward-pointing arrows on the top of each subpanel indicate the $S/N$ of a galaxy with $R_e$ = 5 pixels, $e$ = 0,  and $m_H$ = 20 or 25 (one orbit). Typical red nugget galaxies in CANDELS have $m_H \approx 20-23$.

A notable feature in Figure \ref{fig:3comp1} are the results for galaxies with 6 $<$ $n$ $<$ 20. Note that we do not impose an upper limit on the S\'{e}rsic index; $n$ = 20 is an internal upper limit set by {\tt GALFIT}. It is clear that the fits with $n$ $>$ 6 are generally unreliable. S\'{e}rsic profiles with $n$ $>$ 4 have long tails. In a regime where the surface brightness of the galaxy is close to the sky level, the flux in the long tail can be overestimated due to the inherent degeneracy between a S\'{e}rsic profile with large $n$ and any residual background flux. This leads not only to overestimating the total flux of the galaxy but also its size. On the other hand, for all practical purposes, on can correct for this effect by refitting the galaxies with $n$ $>$ 6 by fixing $n$ to 6. Figure \ref{fig:3comp2} demonstrates the significant improvement in the size determination. Comparing Figures \ref{fig:3comp1} and \ref{fig:3comp2} clarifies that most of the outliers in Figure \ref{fig:3comp1} are galaxies with $n$ $>$ 6. It is worth noting that because the simulations are based on a finite sample of 100 nearby galaxy ``templates,'' rescaled in luminosity and size, the behavior of the scatter is only pseudo-random, which leads to horizontal striations in the scatter plots of Figure \ref{fig:3comp1}.

Figures \ref{fig:3comp1} and \ref{fig:3comp2} (and \ref{fig:3comp3}) show that when the objects are binned by measured S\'{e}rsic index, the scatter does not appear symmetric in the lowest bin in $n$.  One might interpret the result as the fit missing the outer parts of the galaxies.   However, that offset is deceptive because the objects are plotted according to their measured S\'{e}rsic index rather than intrinsic index; one does not know the S\'{e}rsic index for real, multi-component galaxies until after the fit.  Because measured S\'{e}rsic indices are positively correlated in a fit with measured sizes (\citealt{Yoon11}), objects selected to have low measured S\'{e}rsic index typically would have smaller measured sizes.  This is seen in the extreme, comparing Figures \ref{fig:3comp1} and \ref{fig:3comp2}, for the bin where $n$ $>$ 6.  In Figure \ref{fig:3comp2}, the measured sizes become much smaller when $n$ $>$ 6 objects are constrained to $n$ = 6.  The apparent systematic bias is therefore an artifact of the S\'{e}rsic index not being an independent variable in the simulation.  In contrast, when all the objects are grouped together without regard to S\'{e}rsic index, the scatter is more symmetric about the mean (top row). 

Tables 5 and 6 summarize the results of the best-fit multi-component models. On average, across the entire $S/N$ range of interest, the size of the galaxies and their total luminosities are slightly underestimated at very low $S/N$. On the other hand, a small but positive trend is present, as seen in Figure \ref{fig:3comp2}, at higher $S/N$ values. The multi-component galaxies are drawn from CGS elliptical galaxies and the S\'{e}rsic indices of these galaxies should peak around $n$ = 4, as seen in our results ($< 15$\% have $n$ $<$ 2.5, and $< 20$\% have $n > 6$ ). 

There is a weak size dependence in the size determination of the galaxies. The systematic offsets of galaxies with $S/N$ $>$ 50 are smaller for smaller galaxies. Figure \ref{fig:ellipse} indicates that the CoG technique to measure sizes can produce systematics even in idealized galaxies when $R_e$ $<$ 2 pixels. Even after correcting for this effect, the systematic offsets for galaxies with 1 $<$ $R_e$ $<$ 5 pixels are still slightly less than what we find for galaxies with larger sizes. The slight trend and scatter compared to idealized simulations therefore illustrate the fundamental differences due to structural complexities.

Figure \ref{fig:3comp3} and Tables 7 and 8 summarize the results of multi-component models with an observed sky background taken from CANDELS UDS $H_{160}$ images. The trends are similar between images with simulated and observed backgrounds.  As expected, the uncertainties are higher for images with the observed background, especially at lower $S/N$.  

\begin{figure}[t]
  \centering
\includegraphics[width=95mm]{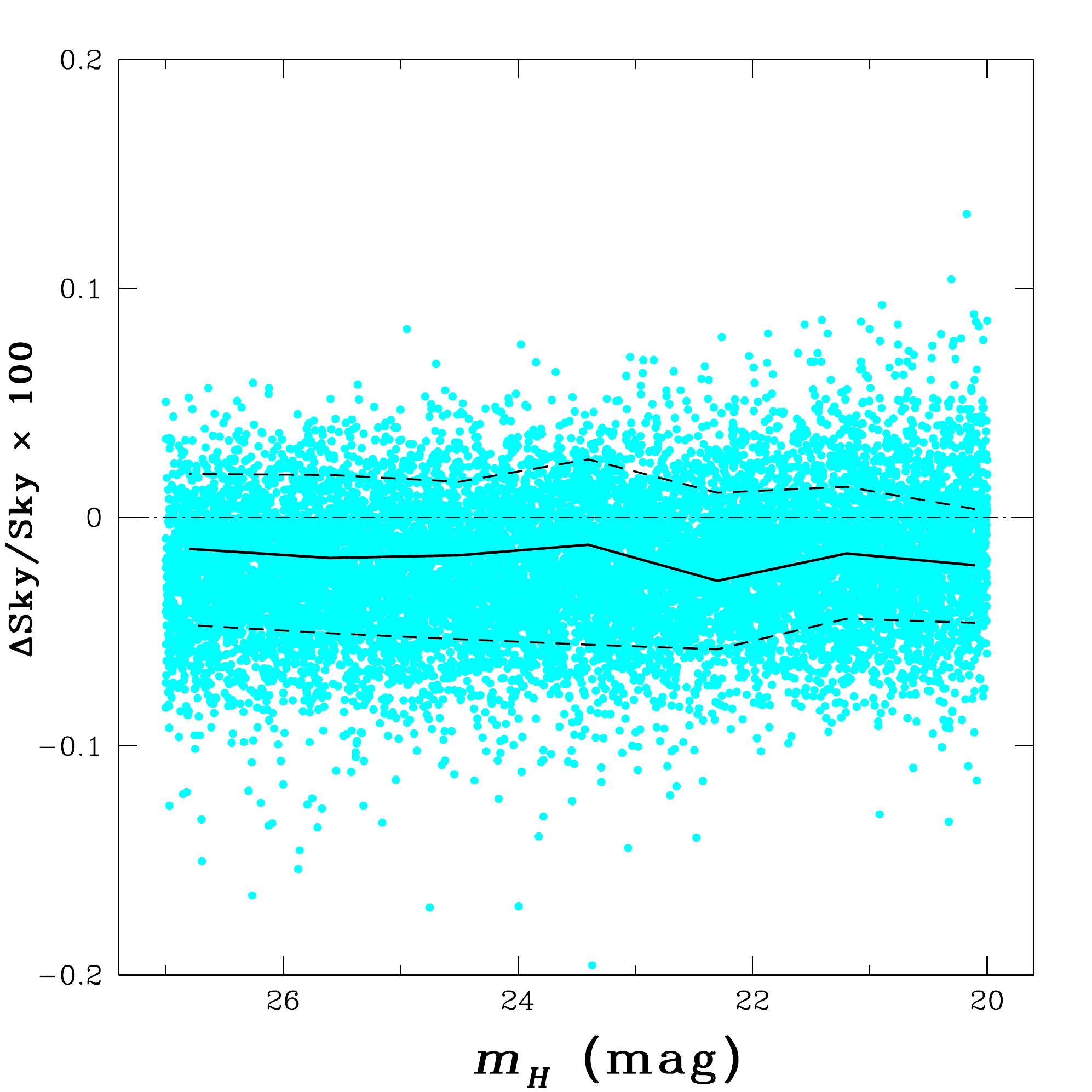}
  \caption{Examining the accuracy of {\tt GALFIT} sky determination. Black solid and dashed lines in the scatter plots indicate the median and 1$\sigma$ uncertainties of different measurements. The dot-dash line shows zero offset. For our simulations, {\tt GALFIT} measures the sky value with accuracy better than a few 0.01$\%$. \label{fig:sky}}
\end{figure}

Under-subtraction or over-subtraction of the sky value can introduce spurious curvature into the brightness profile, especially in the faint, outer regions of the galaxy (e.g., \citealt{MacArthur03}; \citealt{Erwin08}; \citealt{Bernardi10}; \citealt{Yoon11}). Figure \ref{fig:sky} shows the accuracy of {\tt GALFIT} sky determination. It shows that in our simulations {\tt GALFIT} measures the sky value with accuracy better than few 0.01$\%$. Therefore, the systematic errors of single-component fits of multi-component galaxies are not caused by overestimating or underestimating the sky value, but rather merely reflects the limitations of single-component modeling. In fact, even when we fix the sky value to the actual value during the fits, the outcomes do not change.  

The provided tables can be used by future observers to quantify potential systematic offsets and uncertainties in their measurements of structural parameters in high-$z$ galaxies.  Although the main purpose of the simulations is to determine the robustness of the {\tt GALFIT} modeling of high-\emph{z}, compact galaxies, our simulations would be appropriate for galaxies of all redshifts, as long as they fall within our S/N, size, Sersic index, and parameter space. Note, however, that the multi-component simulated galaxies mimic the morphology of only early-type galaxies. A forthcoming work (R. Davari et al., in preparation) will address a broader range of morphological types.

\section{Comparison to Other Studies}

Van der Wel et al. (2012) performed comprehensive single-component simulations based on CANDELS imaging in $H_{160}$, based on their catalog of 6492 objects in GOODS-South. They find that  $m_H$, $R_e$, and $e$ can be inferred with a random accuracy of 20$\%$ or better for galaxies brighter than $m_H$ = 24.5, whereas $n$ can be measured at the same level of accuracy for galaxies brighter than $m_H$ = 23.5 (their Table 3). They conclude that since a typical faint high-\emph{z} galaxy is small and has a low S\'{e}rsic index, 10$\%$-level accuracy in the single-component measurements can be reached down to $m_H\approx 24.5$. For their faintest sources ($m_H$ $>$ 25.5) with large sizes ($R_e$ $>$ 0\farcs4), the uncertainties in the magnitude and structural parameters start to become substantial because they are dominated by the uncertainty in the background estimate. This is in agreement with our single-component results (Tables 1 and 2). We find that at a specific magnitude the scatter is higher for larger galaxies because their surface brightness is lower.

\citet{Trujillo07} simulate 1000 single-component galaxies with properties matching the observed distribution of their objects in $I_{814}$. A background sky, randomly taken from the $I_{814}$ image, is added to the generated galaxies, and the galaxy models are convolved with the observed PSF. Based on their Figure 2, there is no systematic offset for galaxies with $m_I$ $<$ 24.0. For fainter galaxies, only those with $n$ $>$ 2.5, there is less than a 25$\%$ systematic offset toward smaller sizes. They also explore the variation of the PSF within the image to see how it can affect the recovery of the sizes. Using different stars in the image as the PSF, they find that the size estimations are robust to changes in the selected PSF; the scatter is about 10$\%$.  \citet{Abramson13} also use six different PSFs (including an empirical PSF) and find 5$\%$-level accuracy for lower S\'{e}rsic indices and 30$\%$-level accuracy for higher S\'{e}rsic indices. Our Figure \ref{fig:psf3} confirms that the effects are small when using slightly inaccurate PSFs.

However, \citet{Mancini10} claim that for objects with large effective radii and S\'{e}rsic indices, as elliptical galaxies with masses of 2.5 $\times$ $10^{11} M_{\Sun}$ are expected to be, one could substantially underestimate $n$ and $R_e$ for the typical $S/N$ ($\leq$ 50--100) of their sample.  They believe that this reflects the impact of (1 + \emph{z})$^4$ surface brightness dimming of high-$n$ halos. We do not observe any systematic error in size determination of single-component galaxies with $S/N$ $>$ 50 ($m_H$ $<$ 24 for typical \emph{z} = 2 galaxies).

The sizes of single-component galaxies can be underestimated if the dimension of the image is small relative to the galaxy size. This makes the sky determination less reliable.  Size measurements can also be compromised if extended neighbors are not properly masked.  However, when these factors are taken into account, we, in agreement with several previous studies (\citealt{Haussler07}; \citealt{vanDokkum10}; \citealt{Williams10}; Buitrago et al. 2013), believe that the parameters of a single-component galaxy can be retrieved robustly over a wide range of $S/N$.

\citet{Meert13} show that single-component S\'{e}rsic fits of two-component (bulge + disk) simulated galaxies, based on a spectroscopically selected sample from the Sloan Digital Sky Survey (SDSS; \citealt{York00}; \citealt{Stoughton02}), leads to an overestimation of the size (their Figure 8). It holds even for their model galaxies without any noise (their Figure 8$a$). 

\citet{Mosleh13} also simulate two-component galaxies based on an SDSS sample. They examine the robustness of size measurements based on non-parametric fitting as well as single- and two-component S\'{e}rsic fits.  For their modeled galaxies at \emph{z} = 0, they find that their non-parametric and two-component S\'{e}rsic fits provide the most robust $R_e$ measurements, while those based on single-component S\'{e}rsic fitting often overestimate the size, especially for massive red/early-type galaxies (their Figure 3). Redshifting their galaxies to \emph{z} = 1, they find that the single-S\'{e}rsic fitting of two-component galaxies yield reliable size measurements, likely due to the smaller structures being washed out at high redshifts (their Figure 7).  

The two-component simulation results of  \citet{Meert13} and \citet{Mosleh13} qualitatively agree with our results for multi-component galaxies (Figures 6--8). However, \citet{Mosleh13} finds smaller systematic offsets which could be caused by a considerable fraction of disk-dominated galaxies in their sample. Note that our simulations are limited to early-type galaxies.

\section{Implications for Red Nuggets}

One of the main goals of this study is to examine whether or not there is a bias in measuring properties of the red, massive compact galaxies at \emph{z} = 2. These so-called red nuggets are found to have a median stellar mass $M_*$ $\approx$ $10^{11}$ $M_{\Sun}$ and a median effective radius $R_e$ $\leq$ 2.0 kpc (\citealt{vanDokkum08}; \citealt{vanderWel11}). Red nuggets  are found to have $H_{160}$ $<$ 24, with a few having $H_{160}$ = 23--24 (\citealt{Szomoru12}).  They are seen at high $S/N$ ($\geq$ 100, for one {\it HST}\ orbit). 

Assuming that the light distribution of these compact galaxies resembles a S\'{e}rsic profile, Figures \ref{fig:1comp1} and \ref{fig:1comp2} and Tables 1 and 2 show that there is no systematic bias in the size, total luminosity, S\'{e}rsic index, and ellipticity of these galaxies. The uncertainties for measuring $R_e$, $m_H$, $n$, and $e$ are less than 20$\%$, 0.2, 0.2, and 10$\%$, respectively, for galaxies with 23 $<$ $H_{160}$ $<$ 24. The uncertainties are much smaller still for brighter galaxies.  Thus, provided that red nuggets have S\'{e}rsic profiles, single-component fits of these galaxies are very robust.

On the other hand, the observed compact galaxies may have an envelope with low surface brightness that is difficult to recover in detail. If the envelope does not follow a perfect S\'{e}rsic function or the overall galaxy profile changes from small to large radii, it is commonly speculated that the envelope flux and hence the true effective radius may be underestimated (e.g., \citealt{Hopkins09}). We tackle this problem by taking advantage of H13's morphological study of  $\sim$100 nearby massive elliptical galaxies with median mass $M_{\Sun}$ = 1.3 $\times$ $10^{11}$ $M_{\sun}$.  We directly test the null hypothesis that the red nuggets have undergone no morphological evolution and only passive fading since \emph{z} $\approx$ 2.

Figure \ref{fig:3comp2} and Table 5 and 6 show the best-fit results of the single S\'{e}rsic fits of the multi-component galaxies.  With adequate $S/N$ ($\geq$ 100, comparable to red nuggets studied in CANDELS), the presence of extended envelopes in ellipticals actually leads to a slight {\it overestimation}\ of their sizes rather than an underestimation. This holds over a wide range of sizes (i.e., 1 $<$ $R_e$ $<$ 70 pixels). At much lower $S/N$ values (e.g., $\leq$ 50),  the trend reverses but the effect is very modest.   Conclusion: if red nuggets have structures similar to local massive elliptical galaxies, single-component S\'{e}rsic fits of these galaxies do not underestimate their sizes.  The median size of models of nearby ellipticals rescaled to mimic galaxies at \emph{z} = 2 is $R_e \approx$ 0\farcs5, or 4.2 kpc. Considering that the median size of red nuggets is $R_e \approx 1$ kpc, their sizes have increased by a factor of $\sim$4 if they are the progenitors of local massive ellipticals.

Table 6 shows that the total luminosity of the multi-component galaxies can be overestimated as a result of overestimating the size. This leads to systematically higher masses for these compact galaxies. However, this effect is less than 10$\%$, which, compared to other uncertainties involved in mass measurement, is insignificant. Hence, the total luminosity of multi-component galaxies can be measured reliably via single-component S\'{e}rsic fitting.

\section{SUMMARY}

Recent observations of galaxies at \emph{z} = 2 with {\it HST}/WFC3 have revealed a population of red, compact ($R_e \approx 1 - 2$ kpc), and massive ($M_* \approx 10^{11}$ $M_{\Sun}$) galaxies, the so-called red nuggets. We determine the possible biases and uncertainties in the determination of basic structural parameters of these galaxies, with special emphasis on their sizes.  

For this purpose, we perform two sets of simulations: one based on idealized single-component models and the other on the observed properties of local massive ellipticals. For the local analogs, we generate galaxies with multiple components (generally three) with the properties obtained from \citet{Huang13a}, which in turn are rescaled so as to compare with galaxies observed at \emph{z} = 2.0.  We examine the effects of background noise, the accuracy of the PSF model, and the model fitting method.  We analyze the artificial images in a manner similar to that popularly employed in the literature, namely through two-dimensional {\tt GALFIT} modeling of the light distribution using a single-component S\'ersic profile and compare the retrieved size, luminosity, and other structural parameters with input values. 

We find that: 

\begin{itemize}

\item If the sky estimation has been done robustly and the PSF model is relatively accurate, {\tt GALFIT} can retrieve the properties of single-component galaxies at a wide range of magnitudes without introducing any systematic error at high accuracy (Figures \ref{fig:1comp1} and \ref{fig:1comp2} and Tables 1 and 2).

\item Modeling multi-component galaxies with a single S\'{e}rsic component under realistic conditions does not bias the sizes too low; in fact, sizes tend to be slightly overestimated.  This makes the size evolution problem, if anything, even more dramatic.  The apparent compactness of red nuggets is real; it is not the result of missing faint, outer light (Figures \ref{fig:3comp1}--\ref{fig:3comp3} and Table 5).

\item Models with S\'ersic indices larger than 6 have larger uncertainties and can cause significant systematic errors. Refitting these galaxies by fixing the S\'{e}rsic index to $n$ = 6 provides more reliable results (compare Figure \ref{fig:3comp1} with Figure \ref{fig:3comp2}). 

\item We confirm that massive, compact quiescent galaxies at \emph{z} = 2 are a factor of $\sim$4 smaller than their local counterparts.  

\end{itemize}

\acknowledgements
We thank the anonymous referee for a careful reading of the paper and for making detailed suggestions that improved the understanding of this work. RD has been funded by a graduate student fellowship awarded by Carnegie Observatories; we are grateful to director Wendy Freedman for her support.  LCH acknowledges support from the Kavli Foundation, Peking University, the Chinese Academy of Sciences, and the Carnegie Institution for Science. RD would like to thank Prof. G. Canalizo, and Heather L. Worthington for providing long-term support and Minjin Kim, Moein Mosleh, Louis E. Abramson, and Shoubaneh Hemmati for useful discussions. This work was partly funded by NASA grants HST-AR-12818 and HST-GO-12903 from the Space Telescope Science Institute (operated by AURA, Inc., under NASA contract NAS5-26555).  

\appendix

\subsection{The Effects of PSF Accuracy}

\begin{figure}[t]
  \centering
\figurenum{A1}
\includegraphics[width=95mm]{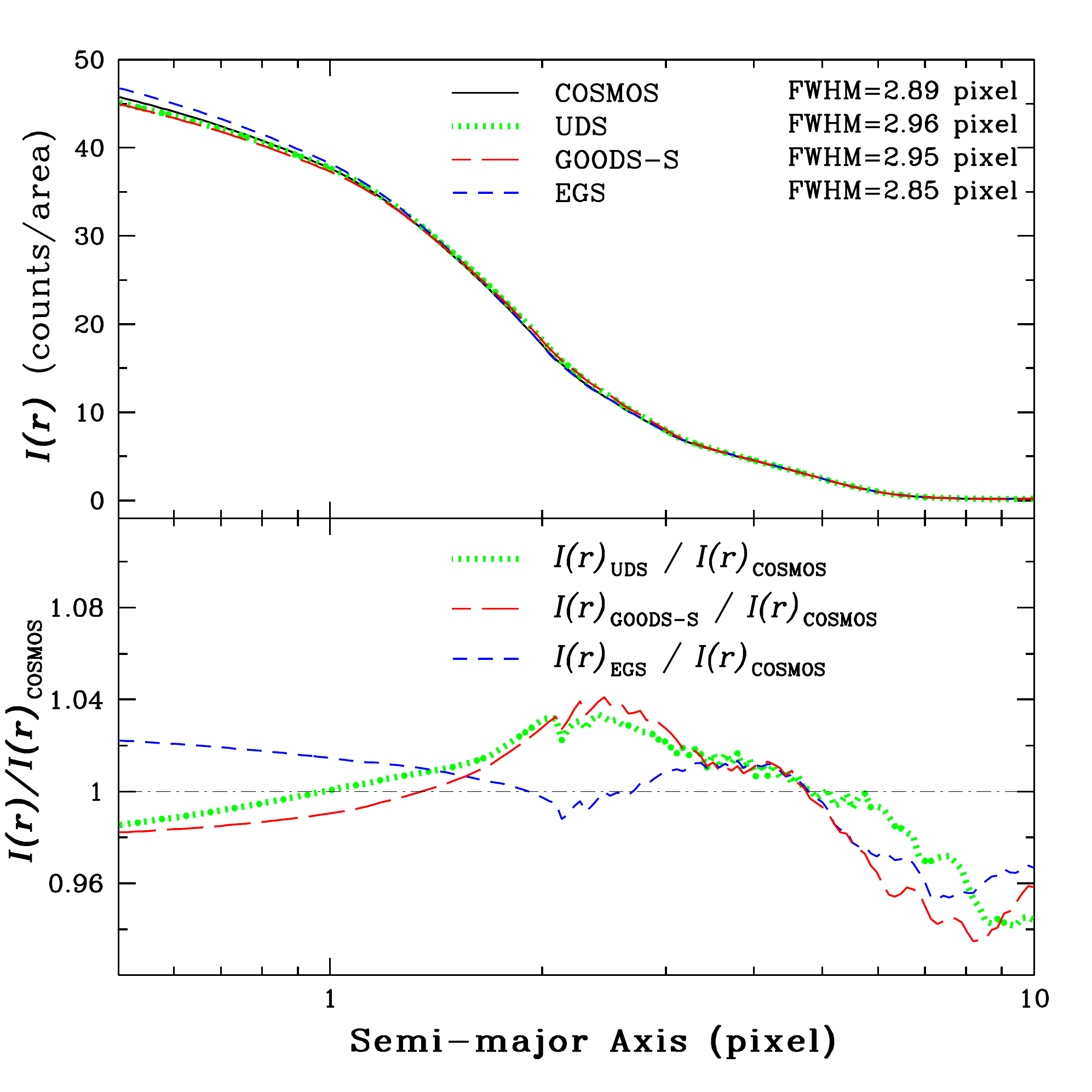}
  \caption{CANDELS $H_{160}$ PSF models for different fields. The PSF models are contructed by using TinyTim for the central and outer parts of the PSF and bright stars for intermediate distances from the center. The top panel shows the one-dimensional light profile of different PSF models. The bottom panel shows the differences between the structure of different PSFs. The dot-dash line marks the case when there is no offset.\label{fig:psf4}}
\end{figure}

An important factor in obtaining a best-fit model is the accuracy of the PSF. The effective radii of high-$z$ galaxies are comparable to and in some cases smaller than the FWHM of the PSF. Therefore, one may expect a considerable offset in {\tt GALFIT} measurements when an inaccurate PSF is used (i.e., a PSF with different FWHM and/or different structure than the PSF used for generating the galaxy model). To study this effect, we first generate about 3,000 single-component galaxies with 0.8 $<$ $R_e$ $<$ 4 pixels, which are convolved with the CANDELS COSMOS $H_{160}$ PSF. Then, we fit each galaxy with the  $H_{160}$ PSF from several major {\it HST}\ surveys (CANDELS, COSMOS, UDS, GOODS-S, and EGS) and compare the results. The PSFs use a combination of a hybrid TinyTim (\citealt{Krist11}) PSF and a stacked empirical stellar PSF.  The motivation for constructing such a hybrid PSF is that the TinyTim PSFs are better in the core region (where centroiding issues tend to broaden empirical PSFs), while empirical PSFs fit real stars better in the wings (\citealt{vanderWel12}). Figure \ref{fig:psf4} shows that these PSF models have different core sizes (i.e. FWHM) and outer halo wing compared to the COSMOS PSF. The PSF in COSMOS has an intermediate core size. 

\begin{figure}
\figurenum{A2}
  \centering
\includegraphics[width=95mm]{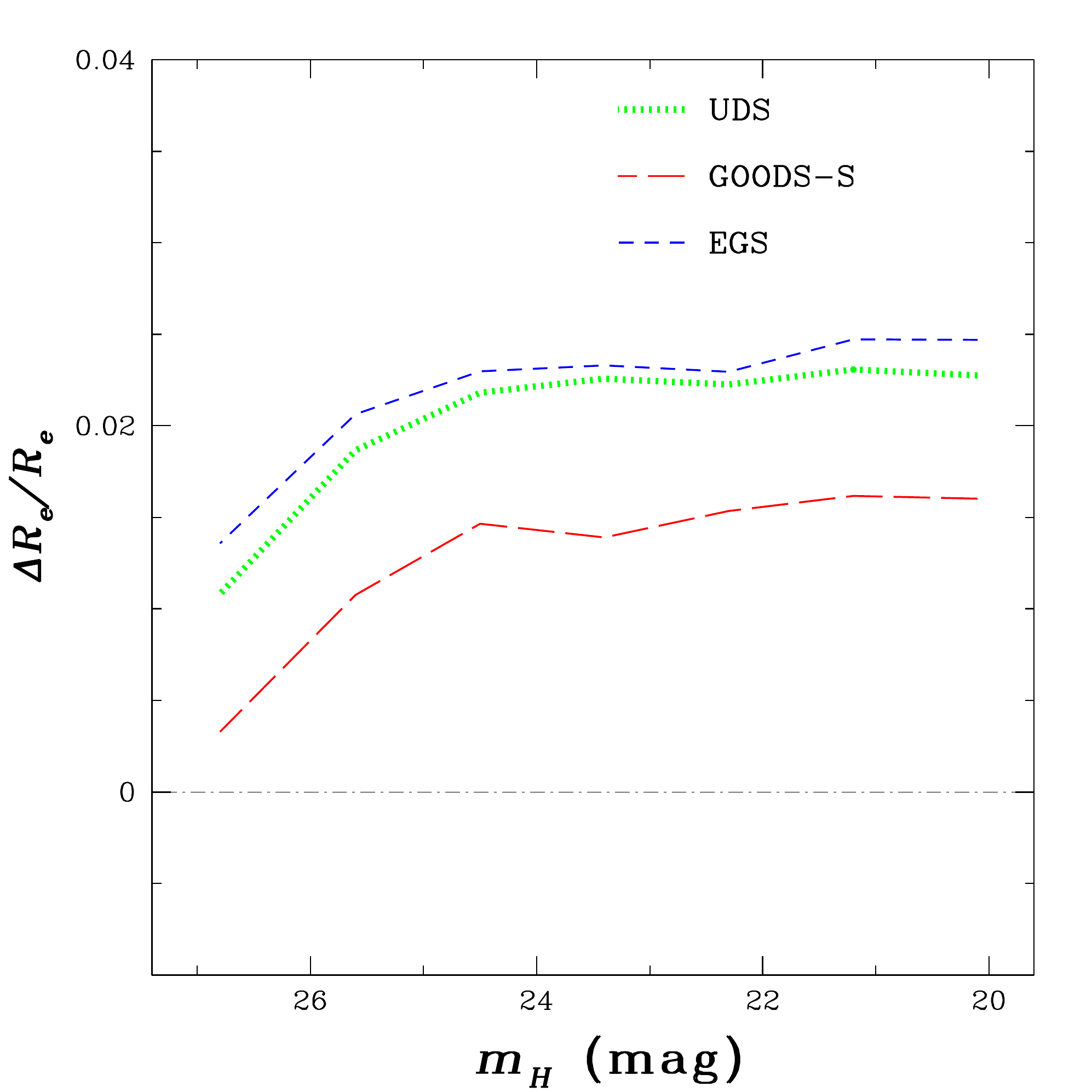}
  \caption{Fitting galaxies with PSFs with different structures and/or FWHMs than the intrinsic PSF of the image leads to biases that are insignificant. The systematic offsets for $R_e$ are less than 2$\%$, and the uncertainties are less than 4$\%$. The dot-dash line shows zero offset. \label{fig:psf3}}
\end{figure}

Figure \ref{fig:psf3} shows that the exact choice of PSF, within the range of realistic PSF models we explored, is not important for the purposes of galaxy size determination.  The systematic offsets are less than 2$\%$ and the uncertainties are less than 4$\%$. The offsets are smaller for best-fit models with lower S\'{e}rsic indices.

\end{document}